\documentclass[aps,prb,twocolumn,shortbibliography,superscriptaddress,article]{revtex4-1}
\usepackage{epsfig}
\usepackage{epstopdf}
\usepackage{amsmath}
\usepackage{amsfonts}
\usepackage{amssymb}
\usepackage{hyperref}
\usepackage{bm}
\usepackage{makecell}
\usepackage{rotating}
\usepackage{hyperref}
\usepackage{multirow}
\usepackage{graphicx}
\usepackage{booktabs}
\usepackage{siunitx}
\usepackage{makecell}
\usepackage{color, soul}

\usepackage{graphicx}
\usepackage{dcolumn}
\usepackage{bm}
\usepackage{color}

\usepackage{tikz,xcolor,hyperref}

\definecolor{lime}{HTML}{A6CE39}
\DeclareRobustCommand{\orcidicon}{%
	\begin{tikzpicture}
	\draw[lime, fill=lime] (0,0)
	circle [radius=0.16]
	node[white] {{\fontfamily{qag}\selectfont \tiny ID}};
	\draw[white, fill=white] (-0.0625,0.095)
	circle [radius=0.007];
	\end{tikzpicture}
	\hspace{-2mm}
}

\foreach \x in {A, ..., Z}{%
	\expandafter\xdef\csname orcid\x\endcsname{\noexpand\href{https://orcid.org/\csname orcidauthor\x\endcsname}{\noexpand\orcidicon}}
}

\begin{document}

\title{Band topology and symmetry-driven magneto-optical response \\
in two-dimensional $d$-wave altermagnets with staggered
spin-orbit coupling}

\author{Meysam Bagheri Tagani\orcidA}
\email{mtagani@magtop.ifpan.edu.pl}
\affiliation{International Research Centre Magtop, Institute of Physics, Polish Academy of Sciences, Aleja Lotnik\'ow 32/46, PL-02668 Warsaw, Poland}
\affiliation{Department of Physics, University of Guilan, P. O. Box 41335-1914, Rasht, Iran}

\author{Carmine Autieri\orcidB}
\email{autieri@magtop.ifpan.edu.pl}
\affiliation{International Research Centre Magtop, Institute of Physics, Polish Academy of Sciences,
Aleja Lotnik\'ow 32/46, PL-02668 Warsaw, Poland}

\author{Wojciech Brzezicki\orcidC}
\affiliation{Institute of Theoretical Physics, Jagiellonian University, ulica S. \L{}ojasiewicza 11, PL-30348 Krak\'ow, Poland}
\affiliation{International Research Centre Magtop, Institute of Physics, Polish Academy of Sciences, Aleja Lotnik\'ow 32/46, PL-02668 Warsaw, Poland}

\date{\today}
\begin{abstract}

Altermagnets combine compensated collinear magnetic order with
momentum-dependent spin splitting, providing a route to transverse
electronic and optical responses without a net ferromagnetic moment.
We develop a strictly periodic four-band tight-binding model for a
two-dimensional $d$-wave altermagnet and distinguish the roles of three
spin--orbit-coupling (SOC) channels: uniform Rashba SOC, a
sublattice-staggered Rashba interaction, and bond-staggered SOC. In the
absence of SOC, the $d$-wave kinetic anisotropy produces spin-polarized
Dirac points on orthogonal Brillouin-zone boundaries, related
by the altermagnetic fourfold spin-group symmetry. Uniform Rashba SOC
mixes the spin sectors and shifts these nodes but preserves the
antiunitary symmetry that forbids an integrated Hall response. The
sublattice-staggered Rashba term breaks this symmetry and activates
transverse optical response, whereas the bond-staggered SOC provides
the mass that gaps the boundary nodes. Their combined action generates
strong Berry-curvature hot spots and, within a narrow parameter window,
an isolated lower two-band manifold with Chern number $C=-2$. For the
representative parameters considered here it is possible to stabilize a Chern insulator phase.
Using covariant-velocity Kubo calculations,
we show that large optical Hall conductivity and circular dichroism
extend well beyond the nonzero-Chern region and are controlled by
SOC-induced avoided crossings and symmetry breaking. We further find
that carrier doping strongly modifies the resonant and dc Hall
responses through Pauli blocking and the occupation of
Berry-curvature hot spots, enabling gate-controlled sign reversals.
These results identify the complementary roles of distinct interfacial
SOC mechanisms in producing topology and tunable magneto-optical
activity in compensated two-dimensional magnets.

\end{abstract}


\maketitle
	
\section{Introduction}

Altermagnetism has emerged as a distinct class of collinear magnetic order that combines features traditionally associated with both antiferromagnets and ferromagnets: a fully compensated sublattice magnetization coexists with large, momentum-dependent spin splitting of electronic bands~\cite{Smejkal2022PRX, Song2025NatRevMat, Mazin2022PRX}. This unconventional magnetic symmetry was identified theoretically as a new paradigm for spintronics, capable of generating spin-polarized transport and optical responses without net magnetization and without the stray fields characteristic of ferromagnets. Recent conceptual developments have established the symmetry foundations of altermagnets and highlighted their potential for efficient spin-current generation, unconventional magnetotransport, and topological phenomena~\cite{Smejkal2022PRX, Bai2024AFM}. These advances have motivated an intense search for material platforms and device concepts that exploit the unique combination of compensated magnetic order and spin-split electronic spectra.

Experimental progress over the past two years has provided compelling evidence for altermagnetic electronic structure. Angle-resolved photoemission spectroscopy (ARPES) and magnetic circular dichroism measurements have directly observed time-reversal-symmetry breaking and momentum-dependent spin splitting in rutile RuO$_2$ and in thin films of CrSb~\cite{Lin2025AdvMat, Reimers2024NatComm, Ding2024PRL, Zhou2025Nature, Yang2025NatComm, Santhosh2025AdvMat, Lu2025NanoLett, Aota2025PRMat, Lin2025AdvMat}. Complementary transport and terahertz spectroscopy studies have revealed anomalous Hall, nonlinear Hall, and ultrafast magneto-optical signatures consistent with altermagnetic order in several compounds and heterostructures~\cite{Lin2025AdvMat, Jiang2025ACSNano, Fang2024PRL, Liu2025PRB, Hu2025NatComm, Gray2024APL, Chen2025PRB, sun2025symmetry}. At the same time, sample-dependent behavior in some materials highlights the sensitivity of altermagnetic order to stoichiometry, strain, and defects, underscoring the need for careful materials engineering and systematic spectroscopic characterization~\cite{Lin2025AdvMat, Chakraborty2024PRB, Karetta2025PRB, Leon2025npjQM}.

Spin--orbit coupling (SOC) plays a dual role in altermagnets. In the weak-SOC limit, altermagnetic spin splitting can arise purely from nonrelativistic exchange and lattice symmetries, but SOC qualitatively reshapes the momentum texture of spin splitting and the Berry curvature that governs anomalous transport and magneto-optical responses~\cite{Hu2025NatComm,  Chen2025PRB,   Leon2025npjQM, Wang2025PRL, PhysRevLett.134.096703,zhou2025contrasting, Autieri2025PRB, Ma2024PRB}. Early theoretical studies often focused on the SOC-free limit to expose the nonrelativistic origin of altermagnetic band splitting; however, neglecting SOC obscures important effects such as SOC-induced gapping of nodal lines, finite Berry-curvature hotspots, and spin-dependent optical selection rules~\cite{Ma2024PRB, Farajollahpour2025npjQM, Vila2025PRB}. More recent models and first-principles works incorporating Rashba or intrinsic SOC have shown that even modest SOC can activate transverse Hall responses, magneto-optical dichroism, and spin–orbit torques that are symmetry-allowed in altermagnets~\cite{Hu2025NatComm, PhysRevLett.134.096703, Han2025PRL, Ko2025arXiv, qu2024extremely, jin2024skyrmion}.

Beyond uniform Rashba or atomic SOC, asymmetric SOC, an SOC term that alternates sign between sublattices or bonds, naturally arises in crystals where local inversion symmetry is broken at the sublattice or bond level, or where orbital hybridization generates bond-dependent spin-dependent hopping phases, as in several rutile and perovskite-derived structures~\cite{Hayami2024Crystals, Browne2021JMCC, Kimura2003PRB, Hogl2020PRL, Zhang2025PRL}. Sublattice-asymmetric SOC has been invoked to explain sublattice-dependent spin–orbit torques, Néel-type spin–orbit control, and band-structure asymmetries that cannot be produced by simple Rashba coupling~\cite{Zhang2025PRL, Smejkal2018Book}. Experimentally, signatures consistent with asymmetric SOC have been inferred from symmetry-selective optical and transport measurements, spin-resolved ARPES patterns requiring bond-dependent spin textures, and spin–torque efficiencies that scale with sublattice symmetry breaking in engineered heterostructures~\cite{Sekh2025arXiv, Vakili2025PRL, Zhou2024arXiv, Eck2022PRB}. Including asymmetric SOC is therefore essential when (i) the crystal sublattices are inequivalent, (ii) spin-dependent hopping is mediated by noncentrosymmetric ligands, or (iii) experiments reveal asymmetric Berry-curvature features that cannot be generated by uniform SOC alone.

These considerations motivate a focused theoretical study of magneto-optical responses in altermagnets that explicitly incorporates both Rashba and asymmetric SOC. First, experimental observations of spectral time-reversal-symmetry breaking and spin splitting establish altermagnets as promising platforms for SOC-mediated optical phenomena. Second, asymmetric SOC is symmetry-allowed in many candidate materials and lifts residual cancellations of Berry curvature, enabling finite transverse optical conductivity and circular dichroism even in the absence of net magnetization. Third, understanding how different SOC textures control optical selection rules, Berry-curvature hotspots, and Hall-like optical signals is essential for both fundamental classification and device applications.

Despite significant progress, existing theoretical studies of altermagnets have almost exclusively focused on uniform (Rashba or atomic) SOC, which preserves a residual $PT$-like symmetry that forces the Berry curvature and optical Hall conductivity to vanish. As a result, Rashba-only $d$-wave altermagnets remain magneto-optically silent even though they exhibit strong momentum-dependent spin splitting. In this work, we identify a distinct and symmetry-allowed SOC channel sublattice-asymmetric spin--orbit coupling that breaks this residual cancellation. We show that staggered SOC is the minimal microscopic ingredient capable of generating finite Berry curvature, optical Hall conductivity, and circular dichroism in a compensated $d$-wave altermagnet, without requiring net magnetization or external magnetic fields. This mechanism is absent in all Rashba-only models and has not been explored in previous magneto-optical studies of altermagnets.

\section{Representative Materials and Model Hamiltonian}


We consider two-dimensional materials based on the Lieb or inverse-Lieb lattice, in which the crystalline $C_4$ rotation connects the two magnetic sublattices and supports a d-wave altermagnetic phase. In the nonrelativistic limit, this magnetic structure produces momentum-dependent spin-polarized band dispersions while maintaining a vanishing net magnetization. Recent theoretical studies have identified monolayer V$_2$Se$_2$O~\cite{SINGH2025101017, liu2025physical}, KV$_2$Se$_2$O~\cite{li2025magnetic}, CrO~\cite{10.1063/5.0147450, li2026manipulating, ch41-gzv1}, and M$_2$WS$_4$ (M=$\mathrm{Mn}$,$\mathrm{Fe}$,$\mathrm{Co}$)~\cite{Xu2025} as representative two-dimensional d-wave altermagnets with the centrosymmetric space group P4/mmm (No.~123). This space group preserves spatial inversion and mirror symmetry with respect to the basal plane.

A second class is formed by inversion-asymmetric or Janus structures, including V$_2$SeTeO~\cite{bezzerga2024giant}, Cr$_2$SeO~\cite{Khan2025}, and Cr$_2$BAl~\cite{Sattigeri2025}. Such systems can crystallize in the noncentrosymmetric space group P4mm (No.~99), which lacks inversion symmetry and the horizontal mirror plane~\cite{ma2021multifunctional, Zhu2023NanoLett}, while retaining the crystalline ingredients required for two-dimensional d-wave altermagnetism.

The distinction between the two structural classes is particularly important for the allowed spin--orbit interactions. In an isolated crystal with the centrosymmetric space group P4/mmm, a uniform Rashba SOC is forbidden by global inversion symmetry. Nevertheless, locally inversion-asymmetric environments can generate sublattice-dependent SOC fields whose signs are opposite on inversion-related magnetic sites. Furthermore, deposition on a substrate generally breaks the equivalence between the two sides of the monolayer and can induce both sublattice-staggered Rashba hopping and bond-dependent spin--orbit hybridization.

In the noncentrosymmetric P4mm class, a uniform Rashba interaction is symmetry allowed even in the freestanding monolayer. A substrate may additionally generate staggered and bond-dependent SOC components by modifying the local crystal fields and hybridization pathways of the two magnetic sublattices. The effective model introduced below therefore contains a uniform inter-sublattice Rashba term together with two substrate-induced contributions: a sublattice-staggered Rashba interaction and a bond-staggered spin-dependent hopping. This formulation allows the same Hamiltonian to describe both globally centrosymmetric and globally noncentrosymmetric material classes by selecting the appropriate SOC parameters.

The altermagnetic Lieb-lattice structure has been experimentally realized or identified in several monolayer compounds~\cite{wu2024valley, kaushal2025altermagnetism, durrnagel2025altermagnetic, wang2025valley,Sattigeri2025}. More recently, a bilayer realization of the Lieb-lattice geometry has been theoretically proposed as a route to enhance the N'eel temperature~\cite{sun2025altermagnetizingfeseliketwodimensionalmaterials}.

In the following, we refer to materials with the centrosymmetric space group P4/mmm as the centrosymmetric material class, represented by CrO, V$_2$Se$_2$O, and KV$_2$Se$_2$O. Materials with the noncentrosymmetric space group P4mm, including Janus monolayers such as V$_2$SeTeO, are referred to as the noncentrosymmetric or Janus material class. Within the effective Hamiltonian, the two classes are distinguished primarily by whether the uniform Rashba coupling is forbidden or allowed in the absence of a substrate, while the substrate-induced staggered SOC terms may be present in either class.

To describe the two material classes introduced above, we employ a minimal tight-binding model for the two magnetic sublattices of a two-dimensional d-wave altermagnet on a Lieb lattice. The two magnetic sites, denoted by A and B, are related by the crystalline $C_{4z}$ rotation and host spinful electronic states. The altermagnetic phase is generated by combining opposite exchange splittings on the two sublattices with a sublattice-dependent $B_{1g}$ hopping anisotropy. This produces momentum-dependent spin splitting while maintaining zero net magnetization in the nonrelativistic limit~\cite{Liu2025PRL}; see Fig.~1(a). The present four-band Hamiltonian should be understood as an effective low-energy description of the two magnetic sublattices, with the nonmagnetic site of the full Lieb lattice integrated out.

The orientation of the Néel vector plays an essential role in determining the magnetic and optical properties of an altermagnet. For an in-plane Néel vector, spin canting can be allowed within the $ab$ plane, whereas the out-of-plane magneto-optical conductivity $\sigma_{xy}$ remains forbidden by the residual magnetic symmetries. For a Néel vector aligned along the z-direction, the two-dimensional d-wave altermagnet on the Lieb lattice realizes a collinear pure altermagnetic state without spin canting~\cite{Sattigeri2025}. More generally, pure altermagnetic phases with finite spin canting can also occur in other magnetic structures~\cite{Fakhredine25b}. In the following, we restrict our analysis to the experimentally relevant configuration in which the Néel vector is parallel to the z- axis~\cite{mcclarty2025observing,Sattigeri2025}. Although many altermagnetic materials are noncollinear~\cite{Autieri2025PRB}, the z-polarized Lieb-lattice altermagnet considered here remains collinear in the absence of relativistic perturbations~\cite{Sattigeri2025}.

For the present model, the absence of spin–orbit coupling allows each spin block to be represented by a real Hamiltonian. Consequently, the Berry curvature vanishes identically in each spin sector. Moreover, the combined crystalline and antiunitary symmetries of the ideal d-wave altermagnet strongly constrain the transverse optical response. To access finite Berry curvature and magneto-optical activity, we introduce substrate-induced spin--orbit coupling. In addition to the conventional inter-sublattice Rashba interaction, we include two symmetry-distinct interface contributions: a sublattice-staggered Rashba hopping and a bond-staggered spin-dependent hybridization. The former lowers the rotational symmetry from $C_4$ to $C_2$ and removes the antiunitary symmetry that forbids the integrated transverse response, while the latter provides a symmetry-allowed mass for the nodal crossings on the Brillouin-zone boundary. The combination of these terms provides a minimal periodic description of substrate-controlled Berry curvature and magneto-optical conductivity.
We take the square-lattice Bravais vectors to be

\[
\mathbf a_x=(1,0),\qquad \mathbf a_y=(0,1),
\]
and choose the intra-unit-cell positions of the two magnetic sites as
\[
\mathbf r_A=\left(\frac12,0\right),\qquad
\mathbf r_B=\left(0,\frac12\right).
\]
The four nearest-neighbor A-to-B bonds are represented by the cell displacements
\[
\mathcal S=
\left\{
0,\mathbf a_x,-\mathbf a_y,\mathbf a_x-\mathbf a_y
\right\}.
\]
For a displacement $\boldsymbol\Delta\in\mathcal S$, the corresponding physical bond vector is
\[
\mathbf d_{\boldsymbol\Delta}
=
\mathbf r_B+\boldsymbol\Delta-\mathbf r_A,
\qquad
\hat{\mathbf d}_{\boldsymbol\Delta}
=
\frac{\mathbf d_{\boldsymbol\Delta}}
{|\mathbf d_{\boldsymbol\Delta}|}.
\]
We introduce the sublattice index
\[
\eta_A=+1,\qquad \eta_B=-1,
\]
or, equivalently, $\eta_\alpha=(-1)^{\alpha+1}$ for $\alpha=1,2$. In terms of the spinors

\[
c_{\alpha,\mathbf R}
=
\begin{pmatrix}
c_{\alpha,\mathbf R,\uparrow}\\
c_{\alpha,\mathbf R,\downarrow}
\end{pmatrix},
\]
the real-space Hamiltonian is written as
\begin{align}
\label{eq:Ham1}
H={}&
t\sum_{\mathbf R,\boldsymbol\Delta\in\mathcal S}
\left(
c_{A,\mathbf R}^{\dagger}
c_{B,\mathbf R+\boldsymbol\Delta}
+\mathrm{h.c.}
\right)
\nonumber\\
&+
\sum_{\alpha,\mathbf R,\mu=x,y}
\left(
t_{\alpha}^{\mu}
c_{\alpha,\mathbf R}^{\dagger}
c_{\alpha,\mathbf R+\mathbf a_\mu}
+\mathrm{h.c.}
\right)
\nonumber\\
&+
M\sum_{\alpha,\mathbf R}
\eta_\alpha\,
c_{\alpha,\mathbf R}^{\dagger}
\sigma_z
c_{\alpha,\mathbf R}
\nonumber\\
&+
\lambda_R
\sum_{\mathbf R,\boldsymbol\Delta\in\mathcal S}
\left[
ic_{A,\mathbf R}^{\dagger}
\left(
\hat{\mathbf d}_{\boldsymbol\Delta}
\times\boldsymbol\sigma
\right)_z
c_{B,\mathbf R+\boldsymbol\Delta}
+\mathrm{h.c.}
\right]
\nonumber\\
&+
\lambda_s
\sum_{\alpha,\mathbf R}
\eta_\alpha
\Big[i
c_{\alpha,\mathbf R}^{\dagger}
\sigma_y
c_{\alpha,\mathbf R+\mathbf a_x}
-
ic_{\alpha,\mathbf R}^{\dagger}
\sigma_x
c_{\alpha,\mathbf R+\mathbf a_y}
+\mathrm{h.c.}
\Big]
\nonumber\\
&+
\lambda_b
\sum_{\mathbf R,\boldsymbol\Delta\in\mathcal S}
\nu_{\boldsymbol\Delta}
\left[i
c_{A,\mathbf R}^{\dagger}
\sigma_z
c_{B,\mathbf R+\boldsymbol\Delta}
+\mathrm{h.c.}
\right].
\end{align}
The first term describes spin-independent nearest-neighbor hopping between the two magnetic sublattices. The second term contains intra-sublattice hopping along the two primitive lattice directions. To separate the isotropic and d-wave components, we parameterize
\begin{equation}
t_\alpha^\mu=t_0+\eta_\alpha s_\mu t_d,
\qquad
s_x=+1,\qquad s_y=-1. \nonumber
\end{equation}
Explicitly,
\begin{align}
t_A^x&=t_0+t_d,&
t_A^y&=t_0-t_d,\nonumber\\
t_B^x&=t_0-t_d,&
t_B^y&=t_0+t_d. \nonumber
\end{align}
The $t_d$ contribution changes sign under a $90^\circ$ rotation and transforms according to the $B_{1g}$ representation of the square-lattice point group. Together with the staggered exchange field $M$, it generates the $d_{x^2-y^2}$-wave spin splitting characteristic of the altermagnetic phase~\cite{Liu2025PRL, tagani2026ferroelectric}.

The fourth term in Eq.~\eqref{eq:Ham1} is the conventional inter-sublattice Rashba SOC with strength $\lambda_R$. This interaction is allowed when global inversion symmetry is broken, as in noncentrosymmetric monolayers or films placed on an inversion-breaking substrate~\cite{Liu2025PRL,Belayadi2024PRB}. We neglect the Kane--Mele SOC considered in Ref.~\cite{tagani2026quantumanomaloushallconductivity} and focus instead on interface-induced Rashba-type and bond-dependent SOC.

The fifth term describes a sublattice-staggered Rashba interaction with strength $\lambda_s$. It corresponds to opposite local Rashba fields on the two magnetic sublattices and may arise when the substrate generates inequivalent local inversion environments at the A and B sites~\cite{Hogl2020PRL,Belayadi2024PRB}. Because this term contains $\sin k_x$ and $\sin k_y$ in momentum space, it represents an intra-sublattice spin-dependent hopping rather than an onsite spin--orbit field.

The final term is a bond-staggered interface SOC with strength $\lambda_b$. Its bond-dependent signs are chosen as
\begin{equation}
\nu_{0}
=
\nu_{\mathbf a_x-\mathbf a_y}
=+1,
\qquad
\nu_{\mathbf a_x}
=
\nu_{-\mathbf a_y}
=-1. \nonumber
\end{equation}
This term phenomenologically describes substrate-assisted spin-dependent hybridization whose sign alternates between the two diagonal bond orientations. It is periodic, time-reversal symmetric in the absence of the exchange field, and generates a mass at the boundary nodal points that are not generically removed by the sublattice-staggered Rashba interaction alone.

To obtain a strictly periodic Bloch Hamiltonian, we use the cell-periodic Fourier convention
\begin{equation}
\label{eq:periodicFT}
c_{\alpha,\mathbf R}
=
\frac{1}{\sqrt{N}}
\sum_{\mathbf k}
e^{i\mathbf k\cdot\mathbf R}
c_{\alpha,\mathbf k}.
\end{equation}
In the basis

\[
\Psi_{\mathbf k}
=
\left(
c_{A,\mathbf k,\uparrow},
c_{A,\mathbf k,\downarrow},
c_{B,\mathbf k,\uparrow},
c_{B,\mathbf k,\downarrow}
\right)^T,
\]

the Hamiltonian takes the block form
\begin{equation}
\label{eq:Hk}
H(\mathbf k)
\begin{pmatrix}
h_A(\mathbf k)&\Gamma(\mathbf k)\\
\Gamma^\dagger(\mathbf k)&h_B(\mathbf k)
\end{pmatrix}.
\end{equation}

The diagonal sublattice blocks are
\begin{align}
h_\alpha(\mathbf k)
={}&
\left[
\varepsilon_+(\mathbf k)
+\eta_\alpha\varepsilon_-(\mathbf k)
\right]\sigma_0
+\eta_\alpha M\sigma_z
\nonumber\\
&+
2\eta_\alpha\lambda_s
\left(
\sin k_y\,\sigma_x
-\sin k_x\,\sigma_y
\right),
\end{align}
where
\begin{align}
\varepsilon_+(\mathbf k)
&=
2t_0\left(\cos k_x+\cos k_y\right),\\
\varepsilon_-(\mathbf k)
&=
2t_d\left(\cos k_x-\cos k_y\right). \nonumber
\end{align}
Equivalently,
\begin{equation}
\varepsilon_\alpha(\mathbf k)
=
2t_\alpha^x\cos k_x+
2t_\alpha^y\cos k_y,
\qquad
\varepsilon_\pm
=
\frac{\varepsilon_A\pm\varepsilon_B}{2}. \nonumber
\end{equation}
The inter-sublattice block is
\begin{equation}
\Gamma(\mathbf k)
=
\Gamma_t(\mathbf k)
+
\Gamma_R(\mathbf k)
+
\Gamma_b(\mathbf k).
\end{equation}
and the spin-independent nearest-neighbor hopping is
\begin{equation}
\Gamma_t(\mathbf k)
=
t
\left(1+e^{ik_x}\right)
\left(1+e^{-ik_y}\right)\sigma_0.
\end{equation}
The conventional Rashba contribution is
\begin{align}
\label{Eq:rashba}
\Gamma_R(\mathbf k)
=
\frac{i\lambda_R}{\sqrt{2}}
\Big[
&
\left(1+e^{ik_x}\right)
\left(e^{-ik_y}-1\right)\sigma_x
\nonumber\\
&+
\left(e^{ik_x}-1\right)
\left(1+e^{-ik_y}\right)\sigma_y
\Big],
\end{align}
while the bond-staggered interface SOC is
\begin{equation}
\Gamma_b(\mathbf k)
=
i\lambda_b
\left(1-e^{ik_x}\right)
\left(1-e^{-ik_y}\right)\sigma_z.
\end{equation}

The equations above constitute a strictly $2\pi$-periodic Bloch Hamiltonian. Indeed, all momentum-dependent coefficients contain only $\sin k_\mu$, $\cos k_\mu$, or integer powers of $e^{\pm ik_\mu}$, and therefore satisfy
\begin{equation}
\label{eq:Gammat}
H(\mathbf k+\mathbf G)=H(\mathbf k) \nonumber
\end{equation} 
for every reciprocal-lattice vector.

The different SOC components play complementary physical roles. The uniform Rashba coupling $\lambda_R$ is associated with global inversion breaking. The local staggered-Rashba coupling $\lambda_s$ produces opposite in-plane SOC fields on the two magnetic sublattices and lowers the rotational symmetry from $C_4$ to $C_2$. In particular, it removes the $C_{4z}\mathcal T$-type constraint that otherwise suppresses the integrated anomalous and magneto-optical Hall responses. The bond-staggered term $\lambda_b$ generates a nonvanishing mass at the Dirac points on the X-M and Y-M Brillouin-zone boundaries. A finite $\lambda_b$ therefore removes the symmetry-unprotected nodal crossings, while $\lambda_s$ allows the resulting Berry curvature to produce a finite transverse optical response.

The two substrate-induced contributions can be parameterized in terms of a single overall interface-SOC strength $\lambda'$ and a mixing angle $\phi$,
\begin{equation}
\lambda_s=\lambda'\cos\phi,
\qquad
\lambda_b=\lambda'\sin\phi. \nonumber
\end{equation}
The angle $\phi$ specifies the microscopic composition of the substrate SOC and is not fixed by symmetry alone. The limit for $\phi=0$ corresponds to a purely sublattice-staggered Rashba interaction, whereas $\phi\neq0$ includes the bond-SOC component required to gap the boundary nodes. In a material-specific description, $\lambda_s$ and $\lambda_b$ may instead be treated as independent parameters or extracted from a Wannier-based tight-binding Hamiltonian.

The staggered exchange field breaks time-reversal symmetry, while the combination of the exchange field and the $B_{1g}$ hopping anisotropy produces the characteristic $d$-wave altermagnetic spin splitting. The uniform Rashba interaction breaks inversion symmetry, whereas the substrate-induced staggered-Rashba term lowers the in-plane point-group symmetry and enables a finite antisymmetric optical conductivity. The resulting Hamiltonian thus incorporates the essential ingredients required to investigate the combined effects of altermagnetism, substrate-induced SOC, nodal-gap formation, Berry curvature, and magneto-optical activity.

Because the Fourier convention in Eq.~\eqref{eq:periodicFT} does not include the intra-unit-cell orbital positions, Berry-curvature and optical-response calculations must retain the corresponding orbital-embedding contribution. The physical velocity operator is
\begin{equation}
v_\mu(\mathbf k)=
\frac{1}{\hbar}
\left[
\frac{\partial H(\mathbf k)}{\partial k_\mu}
+
i\left[
H(\mathbf k),\mathcal R_\mu
\right]
\right],
\end{equation}
where
\begin{equation}
\mathcal R_\mu
=
\begin{pmatrix}
r_{A,\mu}&0\\
0&r_{B,\mu}
\end{pmatrix}
\otimes\sigma_0.
\end{equation}
Using the Hamiltonian and velocity operators in the same Bloch convention guarantees gauge-consistent Berry curvature and optical matrix elements.

In the following, we diagonalize Eq.~\eqref{eq:Hk} numerically and investigate the evolution of the band topology, nodal gaps, Berry-curvature distribution, and frequency-dependent optical conductivity as functions of the altermagnetic and SOC parameters. Unless stated otherwise, we set $t=1$ and use it as the unit of energy.
\section{Results}

\subsection{Symmetry analysis}

We first consider the nonrelativistic limit,
$\lambda_R=\lambda_s=\lambda_b=0$,
in which spin is conserved along the direction of the N'eel vector. For a fixed spin eigenvalue ($s=\pm1$), the Hamiltonian reduces to a two-sublattice block,
\begin{equation}
H_s(\mathbf k)=
\left[\varepsilon_+(\mathbf k)\right]\tau_0
+
H_t(\mathbf k)
+
\left[\varepsilon_-(\mathbf k)+sM\right]\tau_z,
\label{eq:Hspinblock}
\end{equation}
where $H_t(\mathbf k)$ denotes the periodic inter-sublattice hopping generated by $\Gamma_t(\mathbf k)$. Its eigenvalues are
\begin{equation}
\label{eq:nonrelbands}
E_{\nu s}(\mathbf k)=
\varepsilon_+(\mathbf k)
+
\nu
\sqrt{
|\Gamma_t(\mathbf k)|^2+
\left[\varepsilon_-(\mathbf k)+sM\right]^2
},
\quad
\nu=\pm1,
\end{equation}
with
\begin{equation}
|\Gamma_t(\mathbf k)|^2=
16t^2
\cos^2\left(\frac{k_x}{2}\right)
\cos^2\left(\frac{k_y}{2}\right).
\end{equation}
When the sublattice-dependent anisotropy vanishes, $t_d=0$, Eq.~\ref{eq:nonrelbands} becomes independent of (s),
\begin{equation}
E_{\nu,+}(\mathbf k)=E_{\nu,-}(\mathbf k)=
\varepsilon_+(\mathbf k)
+
\nu\sqrt{|\Gamma_t(\mathbf k)|^2+M^2}.
\label{eq:AFMdegeneracy}
\end{equation}
The model therefore describes a conventional collinear antiferromagnet with exact twofold spin degeneracy at every momentum. In an embedded Bloch representation, the nonrelativistic Hamiltonian in this limit is invariant under the combined sublattice and spin transformation, 
$\mathcal U_{\mathrm{AF}}=\tau_x\sigma_x$,
up to the Bloch phase associated with the cell-periodic gauge. This operation exchanges the two magnetic sublattices and reverses $S_z$, mapping every spin-up eigenstate to an orthogonal spin-down eigenstate with the same energy. The degeneracy is therefore an exact symmetry of the minimal model rather than a Kramers degeneracy generated by time reversal alone~\cite{Smejkal2022PRX}.

\begin{figure}
    \centering
    \includegraphics[width=0.99\linewidth]{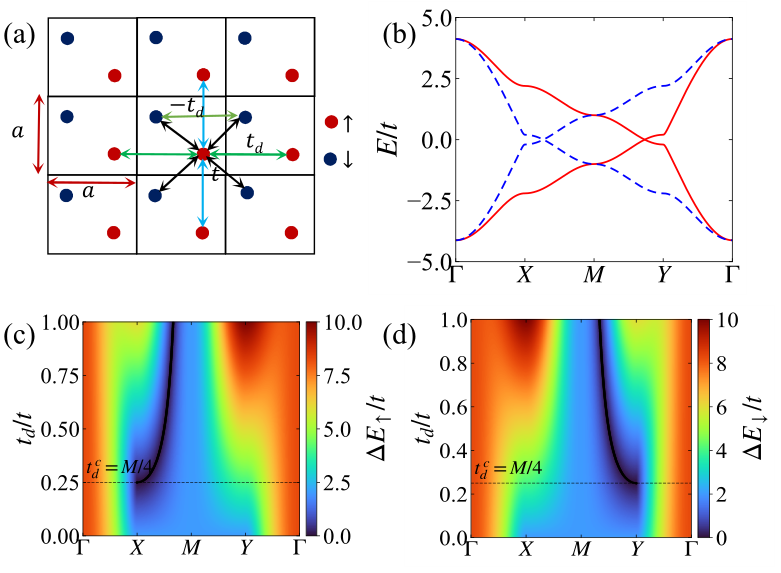}
    \caption{(a) Schematic of the $d$-wave altermagnetic lattice composed of two basis sites. Nearest-neighbor (NN) and next-nearest-neighbor (NNN) hopping processes are indicated. Red (blue) sites indicate spin-up (spin-down) sites. (b) Spin-resolved band structure of the $d$-wave altermagnet along the high-symmetry path $\Gamma$--X--M--Y--$\Gamma$ for parameters $M=1$, $a=1$, $t=1$, and $t_d=0.3$, without SOC terms. Solid (dashed) curves denote spin-down (spin-up) bands. (c)--(d) Evolution of the Dirac points in the Brillouin zone as a function of $t_{d}$ for the spin-up and spin-down channels, respectively. The solid lines trace the momentum positions of the Dirac points for each spin sector.}  
    \label{Fig1}
\end{figure}

In the absence of SOC, the Hamiltonian can also be transformed to a real representation separately in each spin sector. Consequently, the physical Berry curvature vanishes identically, and no intrinsic anomalous Hall or magneto-optical Hall response is generated in this limit.
Altermagnetism emerges for $M\neq0$ and $t_d\neq0$. The d-wave hopping contribution transforms according to the $B_{1g}$ irreducible representation of the square-lattice point group~\cite{Liu2025PRL}. Under a $90^\circ$ rotation,
\begin{equation}
R_{C_4}:(k_x,k_y)\rightarrow(-k_y,k_x), \nonumber
\end{equation}
we have
\begin{equation}
E_{\nu s}\left(R_{C_4}\mathbf k\right)=
E_{\nu,-s}(\mathbf k). \nonumber
\label{eq:C4spinrelation}
\end{equation}
Thus, the $C_{4z}$ rotation connects states with opposite spin polarization. In the nonrelativistic spin-group description, this relation corresponds to a spatial $C_{4z}$ rotation combined with a $\pi$ spin rotation that reverses $ S_z$, which is the characteristic symmetry structure of a d-wave altermagnet~\cite{Smejkal2022PRX,Liu2025PRL}.

The spin splitting of band $\nu$ can be written explicitly as
\begin{align}
\Delta E_\nu(\mathbf k)
&=
E_{\nu,+}(\mathbf k)-E_{\nu,-}(\mathbf k)
\nonumber\\
&=
\nu
\frac{
4M\varepsilon_-(\mathbf k)
}{
\sqrt{
|\Gamma_t|^2+
(\varepsilon_-+M)^2
}
+
\sqrt{
|\Gamma_t|^2+
(\varepsilon_--M)^2
}
}.
\label{eq:AMsplitting}
\end{align}
Equation~\ref{eq:AMsplitting} shows that the spin splitting is proportional to the product $M\varepsilon_-(\mathbf k)$. It therefore changes sign under $C_{4z}$,
\begin{equation}
\Delta E_\nu(R_{C_4}\mathbf k)=
-\Delta E_\nu(\mathbf k), \nonumber
\end{equation}
and vanishes along the d-wave nodal lines
\begin{equation}
\varepsilon_-(\mathbf k)=0
\quad\Longleftrightarrow\quad
\cos k_x=\cos k_y. \nonumber
\label{eq:AMnodallines}
\end{equation}
Equivalently, the spin polarization satisfies
\begin{equation}
\langle S_z(R_{C_4}\mathbf k)\rangle=
-\langle S_z(\mathbf k)\rangle
\end{equation}
for the corresponding $C_4$-related bands. Integration over the full Brillouin zone then gives zero net spin polarization, even though the individual bands are spin split at generic momenta. This alternating spin texture is the defining momentum-space signature of the d-wave altermagnetic state.

Importantly, Eq.~\eqref{eq:AMsplitting} shows that no finite critical value of $t_d$ is required for the onset of altermagnetic spin splitting. Any nonzero $t_d$, together with $M\neq0$, breaks the spin degeneracy away from the nodal lines.  A finite critical value may arise for a separate band inversion, Fermi-surface reconstruction, or topological transition, but not for the existence of altermagnetic spin splitting itself. The restoration of spin degeneracy at vanishing anisotropy is analogous to symmetry-restoring limits discussed in first-principles studies of antiferromagnetic and altermagnetic phases~\cite{Gray2024APL}.

\begin{figure}
    \centering
    \includegraphics[width=0.99\linewidth]{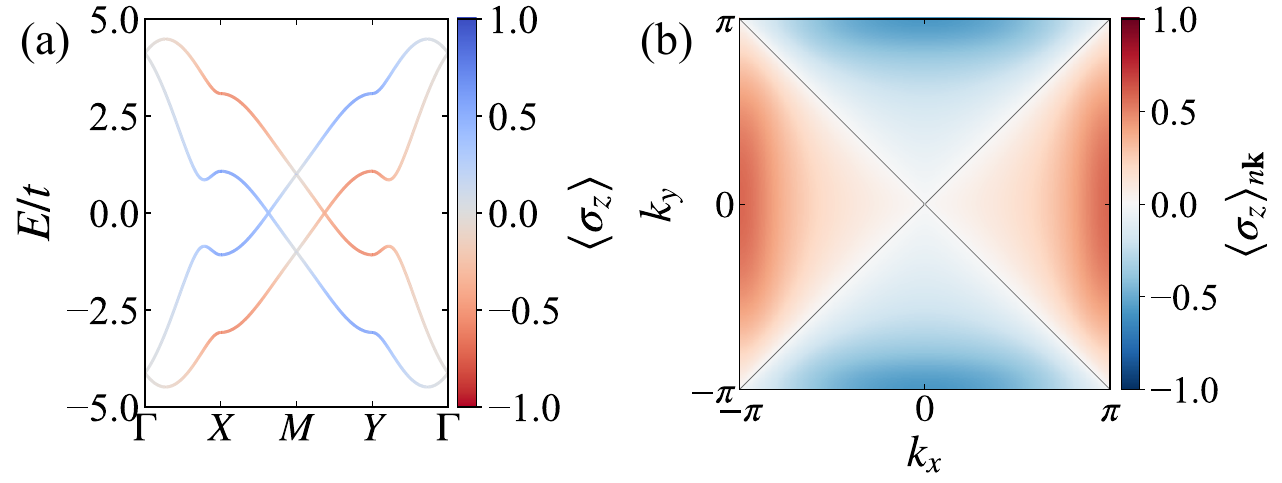}
    \caption{(a) Band structure of the altermagnetic lattice for Rashba spin--orbit coupling $\lambda_R = 0.6t$. All other parameters are identical to those used in Fig.~1. (b) Out-of-plane spin polarization of the low-energy band near the Fermi level in the presence of Rashba SOC. 
    }
    \label{Fig2}
\end{figure}

We next consider the relativistic terms. In the presence of SOC, spin rotations and spatial rotations can no longer be treated independently. For $\lambda_s=0$, the altermagnetic Hamiltonian containing the staggered exchange field, the $B_{1g}$ hopping anisotropy, the uniform Rashba interaction $\lambda_R$, and the bond-staggered SOC $\lambda_b$ remains invariant under the antiunitary operation
$\Theta_4=C_{4z}\mathcal T,$
where $\mathcal T$ is time reversal. The staggered exchange term is odd under $C_{4z}$, because the rotation exchanges the two magnetic sublattices, and is also odd under $\mathcal T$, because time reversal reverses the spin. Their combination therefore leaves the magnetic state invariant. The uniform Rashba and bond-staggered SOC terms are individually time-reversal even and respect the same combined symmetry.

The $\Theta_4$ symmetry imposes an antisymmetric relation on the occupied-band Berry curvature,
\begin{equation}
\Omega_z^{\mathrm{occ}}(\mathbf k)=
-\Omega_z^{\mathrm{occ}}
\left(-R_{C_4}\mathbf k\right).
\label{eq:C4TBerry}
\end{equation}
Although SOC may generate finite local Berry-curvature distributions, Eq.~\eqref{eq:C4TBerry} enforces their cancellation after integration over the Brillouin zone. Consequently,
\begin{equation}
\sigma_{xy}(\omega)=0
\end{equation}
as long as $\Theta_4$ remains a symmetry. In particular, the bond-staggered coupling $\lambda_b$ may gap the Dirac points without, by itself, producing a finite integrated Hall or magneto-optical response.

The sublattice-staggered Rashba interaction,
\begin{equation}
H_s(\mathbf k)=
2\lambda_s\tau_z
\left(
\sin k_y\sigma_x-
\sin k_x\sigma_y
\right),
\end{equation}
has a different symmetry character. It is time-reversal even, but changes sign under the sublattice-exchanging $C_{4z}$ operation. A finite $\lambda_s$ therefore breaks $C_{4z}\mathcal T$ and lowers the rotational symmetry from $C_4$ to $C_2$. Since an out-of-plane Hall response is invariant under $C_{2z}$, the antisymmetric conductivity $\sigma_{xy}(\omega)$ becomes symmetry allowed once $\lambda_s\neq0$, provided that no additional residual mirror or antiunitary symmetry forces it to vanish.

The three SOC terms consequently play distinct roles. The uniform Rashba coupling $\lambda_R$ accounts for global inversion breaking and produces spin mixing between the two sublattices. The bond-staggered coupling $\lambda_b$ generically opens a mass gap at the boundary nodal crossings while preserving the $C_{4z}\mathcal T$ constraint. The sublattice-staggered Rashba coupling $\lambda_s$ removes this constraint and allows the SOC-induced Berry curvature to acquire a nonzero Brillouin-zone integral. The combined presence of altermagnetic spin splitting, a bond-SOC-induced nodal mass, and staggered-Rashba symmetry breaking therefore provides the minimal mechanism for finite anomalous and magneto-optical Hall responses in the present model, despite the absence of conventional ferromagnetic order.

\subsection{Electronic Properties}
Figure~\ref{Fig1}(b) shows the spin-resolved band structure of the
$d$-wave altermagnetic model in the absence of SOC along the
high-symmetry path
$\Gamma$--X--M--Y--$\Gamma$.
The solid red and dashed blue curves denote the spin-down and
spin-up bands, respectively. For the parameters used in
Fig.~\ref{Fig1}(b), two spin-polarized band crossings occur on mutually
orthogonal Brillouin-zone boundaries: the spin-up crossing lies
on the X--M segment, whereas the spin-down crossing lies on the
M--Y segment. These two-dimensional Dirac points are related
by the altermagnetic fourfold spin-group symmetry rather than by
time-reversal or inversion symmetry separately. Consequently,
the spin polarization changes sign under a $90^\circ$ rotation
in momentum space, while its Brillouin-zone average and the net
magnetization remain zero.

In the absence of SOC, $\sigma_z$ is conserved and the full
Hamiltonian separates into two independent spin sectors. For
$\sigma=+1$ (spin-up) and $\sigma=-1$ (spin-down), the two
eigenvalues are expressed as~\eqref{eq:nonrelbands}.
The direct energy difference between the two bands in a fixed
spin sector is therefore
\begin{equation}
\Delta E_{\sigma}(\mathbf{k})
=
E_{\sigma}^{+}(\mathbf{k})
-
E_{\sigma}^{-}(\mathbf{k})
=
2
\sqrt{
\left|\Gamma_t(\mathbf{k})\right|^2
+
\left[
\varepsilon_{-}(\mathbf{k})+\sigma M
\right]^2
}.
\label{eq:spin_direct_gap}
\end{equation}
A Dirac-point crossing requires both contributions under the
square root to vanish:
\begin{align}
\Gamma_t(\mathbf{k})&=0,
\label{eq:nodecondition1}\\ \nonumber
\varepsilon_{-}(\mathbf{k})+\sigma M&=0. \nonumber
\label{eq:nodecondition2}
\end{align}
The first condition is satisfied on the Brillouin-zone
boundaries
$k_x=\pi$, or 
$k_y=\pi$.
Along the X--M boundary, where $k_x=\pi$, one has
\begin{equation}
\varepsilon_{-}(\pi,k_y)
=
-2t_d\left(1+\cos k_y\right). \nonumber
\end{equation}
For the convention $M>0$ and $t_d>0$, the crossing condition can
be satisfied only in the spin-up sector, $\sigma=+1$. Its
position is
\begin{equation}
\mathbf{k}_{W,\uparrow}
=
\left(
\pi,
k_{y,\uparrow}^{W}
\right),
\qquad
k_{y,\uparrow}^{W}
=
\arccos\left(
\frac{M}{2t_d}-1
\right).
\label{eq:weyl_up_position}
\end{equation}
A real solution exists when
$t_d\geq t_d^{W}$, where 
$t_d^{W}=\frac{M}{4}$.
At $t_d=M/4$, the spin-up crossing is created at the X point.
As $t_d$ increases, it moves continuously along X--M toward M,
approaching M only in the limit $t_d/M\rightarrow\infty$.

Similarly, along the M--Y boundary, where $k_y=\pi$,
\begin{equation}
\varepsilon_{-}(k_x,\pi)
=
2t_d\left(1+\cos k_x\right). \nonumber
\end{equation}
The crossing condition is then satisfied in the spin-down
sector, $\sigma=-1$, at
\begin{equation}
\mathbf{k}_{W,\downarrow}
=
\left(
k_{x,\downarrow}^{W},
\pi
\right),
\qquad
k_{x,\downarrow}^{W}
=
\arccos\left(
\frac{M}{2t_d}-1
\right).
\label{eq:weyl_down_position}
\end{equation}
The spin-down crossing is created at Y when $t_d=M/4$ and moves
along Y--M toward M as $t_d$ increases. The spin-up and
spin-down node positions are related by a $90^\circ$ rotation,
$\mathbf{k}_{W,\downarrow}
=
R_{C_4}\mathbf{k}_{W,\uparrow}$,
up to a reciprocal-lattice vector.

Figures~\ref{Fig1}(c) and \ref{Fig1}(d) display
$\Delta E_{\uparrow}(\mathbf{k})$ and
$\Delta E_{\downarrow}(\mathbf{k})$, respectively, as functions
of momentum and $t_d$. The black trajectories follow
Eqs.~\eqref{eq:weyl_up_position} and
\eqref{eq:weyl_down_position}. For $t_d<M/4$, the two bands in
each spin sector remain separated along the selected
high-symmetry path. At $t_d=M/4$, the spin-up and spin-down
crossings appear at X and Y, respectively, and subsequently move
toward M with increasing $t_d$.

The origin of the orthogonal spin-dependent node locations is
the transformation of the $B_{1g}$ hopping term under the
fourfold rotation. The form factor
$\cos k_x-\cos k_y$ changes sign under $C_{4z}$, while the same
rotation exchanges the two magnetic sublattices and therefore
also changes the sign of $\tau_z$. Their product remains
invariant, but the spin polarization of the corresponding band
is reversed. Hence,
\begin{equation}
E_{\sigma}^{\nu}
\left(R_{C_4}\mathbf{k}\right)
=
E_{-\sigma}^{\nu}(\mathbf{k}), \nonumber
\end{equation}
and
\begin{equation}
\left\langle S_z
\left(R_{C_4}\mathbf{k}\right)\right\rangle
=
-
\left\langle S_z(\mathbf{k})\right\rangle. \nonumber
\end{equation}
Importantly, the $d$-wave hopping anisotropy does not break
lattice translation symmetry. Instead, it realizes a
sublattice-dependent kinetic anisotropy that is compatible with
the altermagnetic $C_4$ spin-group operation.

The resulting phase is therefore a spin-polarized
two-dimensional Dirac semimetal: opposite spin sectors host
linearly dispersing crossings on perpendicular Brillouin-zone
boundaries, while the total magnetization remains zero. This
SOC-free altermagnetic spin--momentum locking provides the
starting point for the SOC-induced nodal gaps, Berry curvature,
and magneto-optical responses discussed below.


We first examine the effect of the uniform Rashba spin--orbit
coupling while setting the two substrate-staggered contributions
to zero.
The Rashba interaction is an inter-sublattice spin-dependent
hopping generated by global inversion-symmetry breaking. In the
strictly periodic Bloch representation, it is contained in the
off-diagonal block~\eqref{Eq:rashba}.
Because this term contains the spin matrices \(\sigma_x\) and
\(\sigma_y\), it does not commute with \(\sigma_z\). The spin-up
and spin-down sectors of the nonrelativistic Hamiltonian are
therefore hybridized, and \(S_z\) is no longer an exact quantum
number.

The uniform Rashba interaction preserves the fourfold
crystalline symmetry of the nonmagnetic lattice. In the
altermagnetic state, it also preserves the antiunitary
combination \(C_{4z}\mathcal T\). Consequently, Rashba SOC alone
does not remove the symmetry relating the two orthogonal sets of
boundary nodes and does not make a net transverse Hall response
symmetry allowed. Its principal effects are spin mixing, a shift
of the nodal positions, and a reconstruction of the spin texture.

Although the numerical calculations are carried out using the
strictly periodic Hamiltonian, the energy eigenvalues can be
derived conveniently in the unitarily equivalent embedded
representation. After subtracting the scalar contribution
\(\varepsilon_+(\mathbf{k})\), the Rashba-only Hamiltonian
can be written as
\begin{align}
\widetilde H_R(\mathbf{k})
={}&
\varepsilon_-(\mathbf{k})\tau_z
+
M\tau_z\sigma_z
+
f(\mathbf{k})\tau_x
\nonumber\\
&+
\tau_x
\left[
g_x(\mathbf{k})\sigma_x
+
g_y(\mathbf{k})\sigma_y
\right],
\label{eq:Rashba_embedded_H}
\end{align}
where its  energy eigenvalues are
\begin{equation}
E_{\eta,\zeta}(\mathbf{k})
=
\varepsilon_+(\mathbf{k})-\mu
+
\eta
\sqrt{
S(\mathbf{k})
+
2\zeta\sqrt{Q(\mathbf{k})}
},
\qquad
\eta,\zeta=\pm1,
\label{eq:eigs_Rashba}
\end{equation}
where
\begin{align}
S(\mathbf{k})
={}&
M^2
+
\varepsilon_-^2(\mathbf{k})
+
f^2(\mathbf{k})
+
g^2(\mathbf{k}),
\\ \nonumber
Q(\mathbf{k})
={}&
M^2\varepsilon_-^2(\mathbf{k})
+
\left[
M^2+f^2(\mathbf{k})
\right]
g^2(\mathbf{k}),
\\ \nonumber
g^2(\mathbf{k})
={}&
g_x^2(\mathbf{k})+g_y^2(\mathbf{k}). \nonumber
\label{eq:g2_Rashba}
\end{align}

At half filling and for \(t_0=0\), a crossing between the
two central bands requires
\begin{equation}
S(\mathbf{k}_W)
-
2\sqrt{Q(\mathbf{k}_W)}
=0. \nonumber
\label{eq:Rashba_gapclosing_general}
\end{equation}
The original altermagnetic nodes lie on the Brillouin-zone
boundaries, where the spin-independent inter-sublattice hopping
vanishes. For \(f(\mathbf{k})=0\), the determinant of
\(\widetilde H_R\) reduces to
\begin{equation}
\det\widetilde H_R
=
\left[
\varepsilon_-^2(\mathbf{k})
+
g^2(\mathbf{k})
-
M^2
\right]^2. 
\label{eq:Rashba_boundary_det}
\end{equation}
The boundary nodes therefore satisfy
\begin{equation}
\varepsilon_-^2(\mathbf{k}_W)
+
g^2(\mathbf{k}_W)
=
M^2. \nonumber
\label{eq:Rashba_boundary_condition}
\end{equation}
Thus, uniform Rashba SOC does not automatically generate a mass
at the Dirac points. Instead, it modifies the nodal condition
and shifts the crossings along the Brillouin-zone boundary.

Figure~\ref{Fig2}(a) shows the band structure in the presence of uniform
Rashba SOC. In contrast to Fig.~\ref{Fig1}(b), the individual branches can
no longer be assigned exact spin-up or spin-down labels. The
bands are therefore characterized by their out-of-plane spin
expectation value,
\begin{equation}
s_{n,z}(\mathbf{k})
=
\left\langle
u_{n\mathbf{k}}
\left|
\tau_0\otimes\sigma_z
\right|
u_{n\mathbf{k}}
\right\rangle. \nonumber
\label{eq:spin_expectation}
\end{equation}
The Rashba interaction hybridizes the two spin sectors and
reduces \(\left|s_{n,z}\right|\) below unity at generic momenta.
Nevertheless, the crossings on X--M and M--Y remain visible,
confirming the analytical condition
Eq.~\eqref{eq:Rashba_boundary_det}.

Figure~\ref{Fig2}(b) displays \(s_{n,z}(\mathbf{k})\) for the low-energy
band immediately below the Fermi level. The texture retains the
$d$-wave altermagnetic structure: regions related by a
\(90^\circ\) rotation carry opposite out-of-plane spin
polarization. More precisely, the surviving antiunitary
\(C_{4z}\mathcal T\) symmetry imposes
\begin{equation}
s_{n,z}
\left(
-R_{C_4}\mathbf{k}
\right)
=
-
s_{n,z}(\mathbf{k}), \nonumber
\label{eq:Rashba_spin_symmetry}
\end{equation}
for symmetry-related nondegenerate states. The resulting texture
contains four momentum-space sectors with alternating signs of
the out-of-plane spin polarization, while its Brillouin-zone
average remains zero.


\begin{figure}
    \centering
    \includegraphics[width=0.99\linewidth]{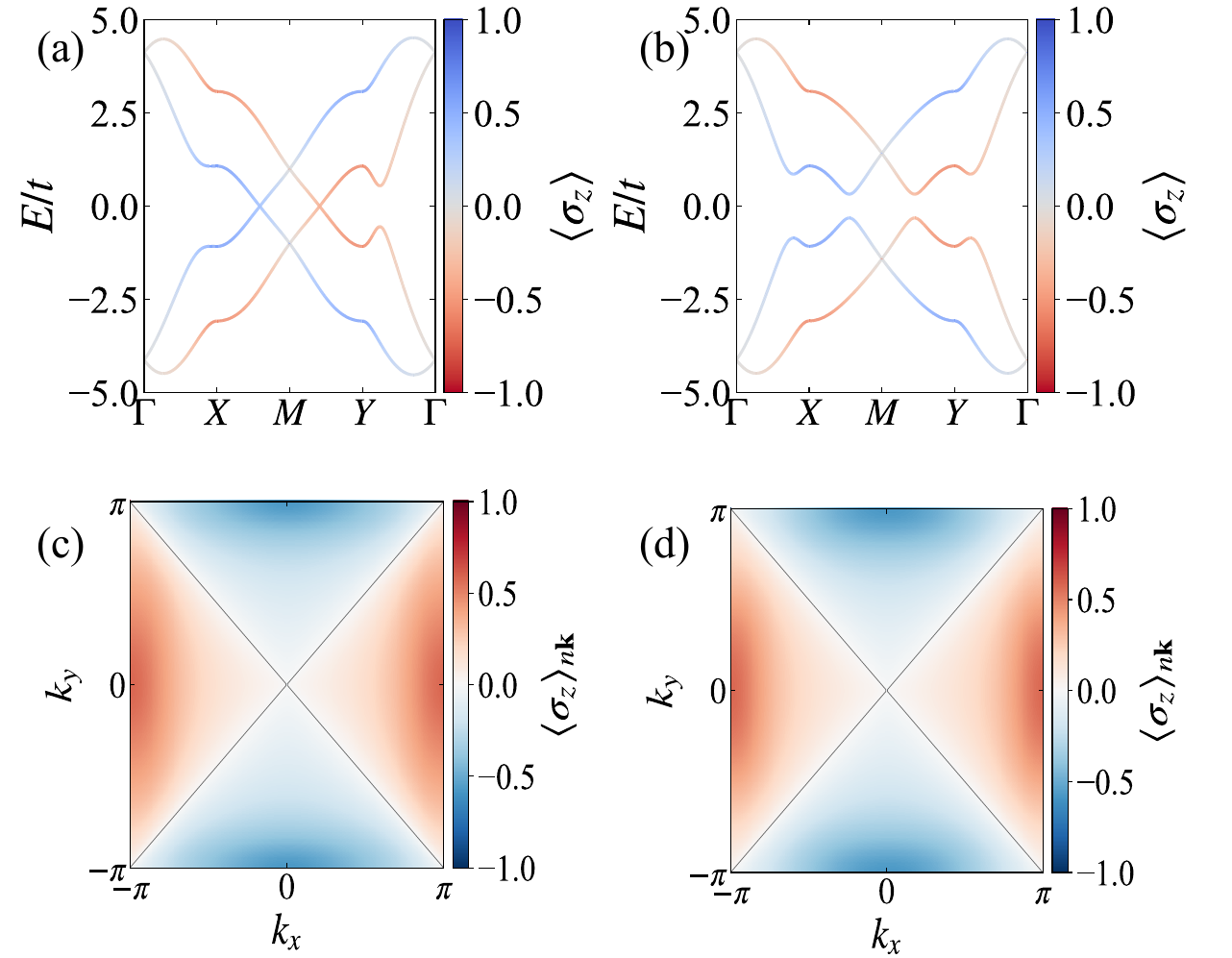}
    \caption{ Band structure of the altermagnetic lattice for (a) staggered Rashba interaction $\lambda_s = 0.25t$, and (b) bond staggered SOC interaction $\lambda_b = 0.25t$. $\lambda_R=0.6t$, and all other parameters are identical to those used in Fig.~1. (c)-(d) Out-of-plane spin polarization of the low-energy band near the Fermi level for (a) and (b), respectively. 
    }
    \label{Fig3}
\end{figure}

We next investigate separately the two substrate-induced SOC
contributions introduced in Eq.~\eqref{eq:Ham1}. This separation
is useful because the sublattice-staggered Rashba interaction
$\lambda_s$ and the bond-staggered SOC $\lambda_b$ have
qualitatively different effects on the nodal spectrum. The former
mixes the two spin sectors and modifies the spin texture, whereas
the latter preserves $S_z$ and acts as a mass term for the
Dirac points.

We first set
\begin{equation}
\lambda_R=\lambda_b=0,
\qquad
\lambda_s\neq0. \nonumber
\end{equation}
The corresponding momentum-space Hamiltonian is
\begin{align}
H_s(\mathbf{k})
={}&
\left[\varepsilon_+(\mathbf{k})\right]\tau_0\sigma_0
+
H_t(\mathbf{k})
+
\varepsilon_-(\mathbf{k})\tau_z\sigma_0
+
M\tau_z\sigma_z
\nonumber\\
&+
2\lambda_s\tau_z
\left(
\sin k_y\,\sigma_x
-
\sin k_x\,\sigma_y
\right),
\label{eq:H_lambda_s}
\end{align}
where $H_t(\mathbf{k})$ denotes the periodic inter-sublattice
hopping associated with $\Gamma_t(\mathbf{k})$.

It is convenient to introduce the effective spin field
\begin{equation}
\mathbf{d}_s(\mathbf{k})
=
\left(
2\lambda_s\sin k_y,
-2\lambda_s\sin k_x,
M
\right)
\end{equation}
and its magnitude
\begin{equation}
D_s(\mathbf{k})
=
\left|\mathbf{d}_s(\mathbf{k})\right|
=
\sqrt{
M^2
+
4\lambda_s^2
\left(
\sin^2 k_x+\sin^2 k_y
\right)
}.
\label{eq:Ds}
\end{equation}
At each momentum, the operator
$\mathbf{d}_s(\mathbf{k})\cdot\boldsymbol{\sigma}$ commutes with
the Hamiltonian. The eigenstates may therefore be labeled by the
helicity index $\zeta=\pm1$, defined through
\begin{equation}
\mathbf{d}_s(\mathbf{k})\cdot\boldsymbol{\sigma}
\,
|u_{\zeta}\rangle
=
\zeta D_s(\mathbf{k})
|u_{\zeta}\rangle. \nonumber
\end{equation}
The four energy branches are
\begin{equation}
E_{\eta,\zeta}^{(s)}(\mathbf{k})
=
\varepsilon_+(\mathbf{k})
+
\eta
\sqrt{
\left|\Gamma_t(\mathbf{k})\right|^2
+
\left[
\varepsilon_-(\mathbf{k})
+
\zeta D_s(\mathbf{k})
\right]^2
}.
\label{eq:eigs_lambda_s}
\end{equation}
Equation~\eqref{eq:eigs_lambda_s} shows that the staggered
Rashba interaction removes conservation of $\sigma_z$, but it
does not generically produce a mass at the original boundary
nodes. 


The spin expectation value can also be obtained analytically.
For a state with helicity $\zeta$,
\begin{equation}
\left\langle
\boldsymbol{\sigma}
\right\rangle_{\zeta}
=
\zeta
\frac{
\mathbf{d}_s(\mathbf{k})
}{
D_s(\mathbf{k})
}.
\label{eq:spin_lambda_s_vector} \nonumber
\end{equation}
In particular,
\begin{equation}
\left\langle
\sigma_z
\right\rangle_{\zeta}
=
\zeta
\frac{M}{D_s(\mathbf{k})}.
\label{eq:spin_lambda_s_z} \nonumber
\end{equation}
Away from degeneracy lines, the highest occupied band corresponds
to the helicity
\begin{equation}
\zeta_{\mathrm{occ}}
=
-\operatorname{sgn}
\left[
\varepsilon_-(\mathbf{k})
\right]. \nonumber
\end{equation}
Its out-of-plane spin polarization is thus
\begin{equation}
s_z^{(s)}(\mathbf{k})
=
-
\operatorname{sgn}
\left[
\varepsilon_-(\mathbf{k})
\right]
\frac{M}{
\sqrt{
M^2+
4\lambda_s^2
\left(
\sin^2 k_x+\sin^2 k_y
\right)
}
}.
\label{eq:occupied_spin_lambda_s}
\end{equation}
Equation~\eqref{eq:occupied_spin_lambda_s} retains the
$d$-wave sign structure of the altermagnetic state, but its
magnitude is reduced below unity because the staggered Rashba
interaction rotates the spin away from the N\'eel axis.

Figure~\ref{Fig3}(a) shows the corresponding band structure. The
Dirac points remain present, although their
positions are displaced relative to the SOC-free case. Since
$\sigma_z$ is no longer conserved, the four branches are colored
according to their expectation value
$\langle\sigma_z\rangle$ rather than labeled as pure spin-up and
spin-down bands. Figure~~\ref{Fig3}(c) displays the out-of-plane spin
texture of the highest occupied band. The alternating four-lobe
pattern survives, while the reduced magnitude of
$\langle\sigma_z\rangle$ reflects the finite in-plane spin
components induced by $\lambda_s$.

Along the lines
$\varepsilon_-(\mathbf{k})=0,$
the two helicity branches become degenerate. The spin expectation
value assigned to an individual eigenvector is then basis
dependent. These points should therefore be masked or treated by
diagonalizing the projected spin operator within the degenerate
subspace when plotting Fig.~~\ref{Fig3}(c).

We next consider the complementary limit
$\lambda_R=\lambda_s=0,~
\lambda_b\neq0.$
In the embedded representation, the bond-staggered SOC takes the
form
\begin{equation}
H_b(\mathbf{k})
=
b(\mathbf{k})\tau_y\sigma_z,
\label{eq:Hb_embedded} \nonumber
\end{equation}
with
\begin{equation}
b(\mathbf{k})
=
-4\lambda_b
\sin\left(\frac{k_x}{2}\right)
\sin\left(\frac{k_y}{2}\right).
\label{eq:b_formfactor} \nonumber
\end{equation}

In contrast to the staggered Rashba term, the bond-staggered SOC
commutes with $\sigma_z$. The two spin sectors therefore remain
independent. For $\sigma=\pm1$, the effective two-sublattice
Hamiltonian is
\begin{align}
H_{\sigma}^{(b)}(\mathbf{k})
={}&
\left[
\varepsilon_+(\mathbf{k})-\mu
\right]\tau_0
+
f(\mathbf{k})\tau_x
+
\sigma b(\mathbf{k})\tau_y
\nonumber\\ 
&+
\left[
\varepsilon_-(\mathbf{k})+\sigma M
\right]\tau_z,
\label{eq:H_sigma_lambda_b}
\end{align}
and its eigenvalues are
\begin{equation}
E_{\eta,\sigma}^{(b)}(\mathbf{k})
=
\varepsilon_+(\mathbf{k})-\mu
+
\eta
\sqrt{
\left|\Gamma_t(\mathbf{k})\right|^2
+
b^2(\mathbf{k})
+
\left[
\varepsilon_-(\mathbf{k})+\sigma M
\right]^2
}.
\label{eq:eigs_lambda_b}
\end{equation}

A zero-energy crossing at half filling requires the simultaneous
conditions
\begin{align}
\Gamma_t(\mathbf{k}_W)&=0,
\\ \nonumber
b(\mathbf{k}_W)&=0,
\\ \nonumber
\varepsilon_-(\mathbf{k}_W)+\sigma M&=0. \nonumber
\label{eq:lambda_b_node_conditions}
\end{align}
The first condition places the crossing on $k_x=\pi$ or
$k_y=\pi$, whereas the second requires $k_x=0$ or $k_y=0$
modulo reciprocal-lattice vectors. Their only common solutions
in the first Brillouin zone are X and Y. At these points, the
remaining condition is satisfied only at the fine-tuned value
$t_d=\frac{M}{4}$.
Thus, for a generic value $t_d\neq M/4$, any finite
$\lambda_b$ removes the band crossings.

Figure~~\ref{Fig3}(b) shows the band structure for finite $\lambda_b$.
Unlike the staggered-Rashba case, the crossings on X--M and M--Y
are replaced by avoided crossings, and a finite direct gap opens
between the second and third bands. For $t_0=\mu=0$, the spectrum
is particle-hole symmetric, so a positive direct gap also
corresponds to an insulating gap at half filling.

Because $\sigma_z$ remains conserved, every nondegenerate
eigenstate satisfies
$\left\langle\sigma_z\right\rangle=\sigma=\pm1$.
For the highest occupied band, the spin sector with the smaller
positive square root in Eq.~\eqref{eq:eigs_lambda_b} is selected,
giving
\begin{equation}
s_z^{(b)}(\mathbf{k})
=
-
\operatorname{sgn}
\left[
\varepsilon_-(\mathbf{k})
\right]
\label{eq:occupied_spin_lambda_b} \nonumber
\end{equation}
away from spin-degenerate lines. Figure~\ref{Fig3}(d) therefore exhibits
a saturated $d$-wave spin texture with
$\langle\sigma_z\rangle=\pm1$. In contrast to Fig.~3(c), the bond
SOC does not cant the spins into the plane; it only modifies the
sublattice composition and opens the nodal gap.

The bond-staggered SOC is time-reversal even and preserves the
$C_{4z}\mathcal T$ symmetry of the altermagnetic state. It can
therefore open a full spectral gap without, by itself, allowing a
nonzero integrated anomalous or magneto-optical Hall response.
The separate calculations in Fig.~\ref{Fig3} reveal the complementary
roles of the two substrate-induced SOC terms: $\lambda_s$ mixes
spin and removes the $C_{4z}\mathcal T$ constraint but leaves the
system nodal, whereas $\lambda_b$ opens the nodal gap while
preserving that constraint. Their simultaneous presence is
therefore required to obtain both a fully gapped spectrum and a
symmetry-allowed transverse magneto-optical response.
\begin{figure}
    \centering
    \includegraphics[width=0.99\linewidth]{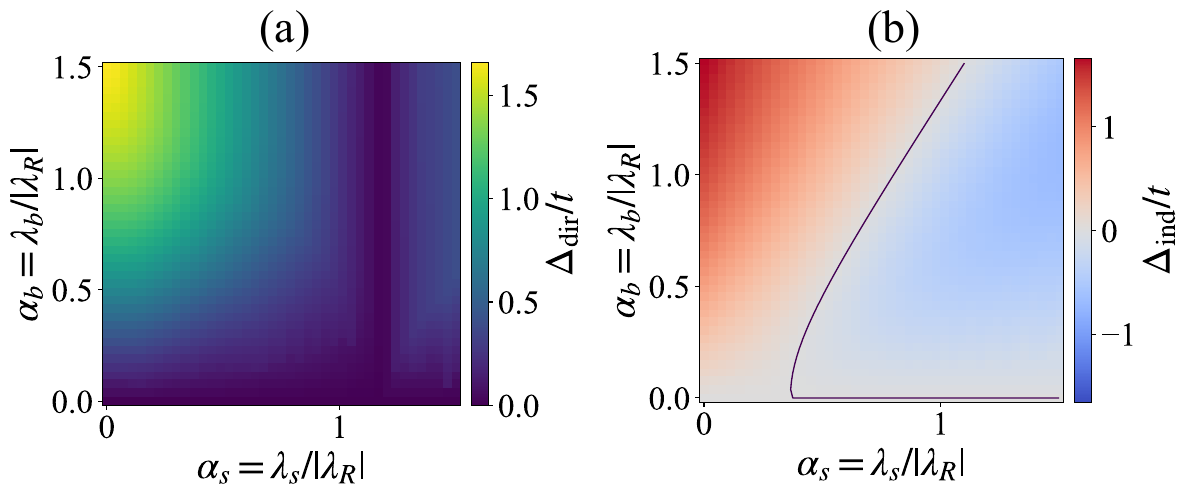}
    \caption{ (a) Minimum direct gap $\Delta_{\mathrm{dir}}/t$ and
(b) indirect gap $\Delta_{\mathrm{ind}}/t$ as functions of the
dimensionless staggered SOC parameters
$\alpha_s=\lambda_s/|\lambda_R|$ and
$\alpha_b=\lambda_b/|\lambda_R|$.
The solid line in panel (b) denotes
$\Delta_{\mathrm{ind}}=0$ and separates the insulating region
from the energy-overlapping metallic regime. $\lambda_R=0.6t$ and other parameters are the same as in Fig. 1.
    }
    \label{Fig4}
\end{figure}

Figure~\ref{Fig4} summarizes the evolution of the half-filling band gaps
as functions of the dimensionless staggered SOC strengths
$\alpha_s=\frac{\lambda_s}{|\lambda_R|}$, and
$\alpha_b=\frac{\lambda_b}{|\lambda_R|}$
with the remaining model parameters held fixed. 

As shown in Fig.~\ref{Fig4}(a), the direct gap remains nearly zero along
the $\alpha_b=0$ axis, consistent with the fact that the
sublattice-staggered Rashba term $\lambda_s$ shifts and
reconstructs the Dirac points but does not generically
gap them. A finite bond-staggered coupling $\lambda_b$, by
contrast, rapidly opens a direct gap. The largest gaps occur for
strong $\alpha_b$ and relatively weak $\alpha_s$, whereas
increasing $\alpha_s$ generally reduces the gap and produces
narrow gap minima associated with further band reconstruction.

The indirect-gap map in Fig.~\ref{Fig4}(b) reveals that a finite direct
gap does not necessarily imply an insulating state. For
sufficiently large $\alpha_b$ and moderate $\alpha_s$,
$\Delta_{\mathrm{ind}}>0$, and the system is a genuine
half-filled insulator. Increasing $\alpha_s$ eventually drives
the indirect gap through zero, as indicated by the solid contour,
and produces an energy-overlapping, direct-gapped metallic
regime with $\Delta_{\mathrm{dir}}>0$ but
$\Delta_{\mathrm{ind}}<0$. These results demonstrate that
$\lambda_b$ provides the primary nodal mass, while $\lambda_s$
controls the subsequent band rearrangement and the stability of
the global insulating gap. The positive-indirect-gap region
therefore defines the physically relevant parameter window for
the topological analysis below.

\begin{figure}
    \centering
    \includegraphics[width=0.99\linewidth]{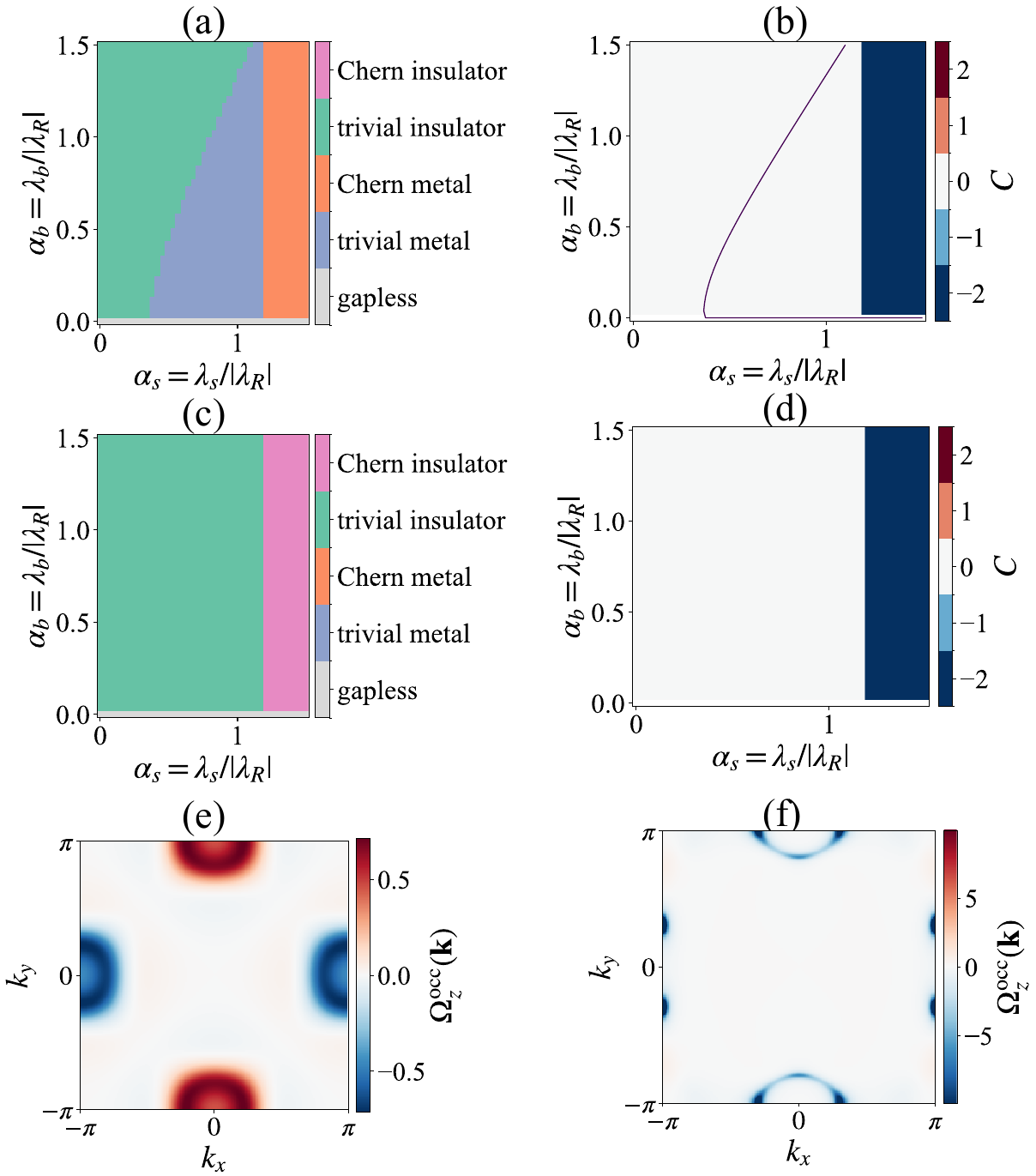}
    \caption{ (a) Spectral and topological phase classification in the
$(\alpha_s,\alpha_b)$ plane.
(b) Chern number of the isolated two-band manifold below the
direct gap. The nontrivial region carries $C=-2$ but lies in the
negative-indirect-gap sector. $M=1t$ is considered. (c) Spectral and topological phase classification, and 
(d) Chern number for $M=2t$. The nontrivial region carries $C=-2$ and lies in the
Chern insulator region, confirming QAHE. 
(e) Occupied-subspace Berry curvature for the trivial gapped
phase at $\alpha_s=0$ and $\alpha_b=1.5$, where the local
contributions cancel and $C=0$.
(f) Berry curvature for $\alpha_s=1.5$ and $\alpha_b=0.56$.
The direct-gapped manifold has $C=-2$, while the negative
indirect gap identifies the state as a Chern band metal rather
than a Chern insulator. $\lambda_R=0.6t$ and other parameters are the same as Fig.1.}
    \label{Fig5}
\end{figure}

\subsection{Berry curvature and Chern-band topology}

Figure~\ref{Fig5} summarizes the spectral and topological character of
the model in the $(\alpha_s,\alpha_b)$ plane. Panel~\ref{Fig5}(a)
distinguishes gapless phases, direct-gapped metals, and
insulating phases, while panel~5(b) shows the Chern number of
the isolated two-band manifold below the direct gap. Most of the
parameter space is topologically trivial, with $C=0$, whereas a
narrow nontrivial region with $C=-2$ appears at large
$\alpha_s$ and intermediate $\alpha_b$.

This restricted topological window reflects the complementary
roles of the two substrate-induced SOC terms. The bond SOC
$\lambda_b$ opens the gaps at the altermagnetic boundary nodes,
but when it acts alone it preserves the symmetry that relates
opposite Berry-curvature sectors. The resulting local curvature
therefore cancels upon integration over the Brillouin zone.
Conversely, the staggered Rashba term $\lambda_s$ removes this
cancellation by breaking the corresponding fourfold
antiunitary constraint, but it does not by itself provide a
robust nodal mass. A nonzero Chern number consequently emerges
only when $\lambda_s$ is sufficiently strong to reorganize the
SOC-gapped bands while $\lambda_b$ remains large enough to keep
the lower two-band subspace directly separated from the upper
bands.

\begin{figure}
    \centering
    \includegraphics[width=1.0\linewidth]{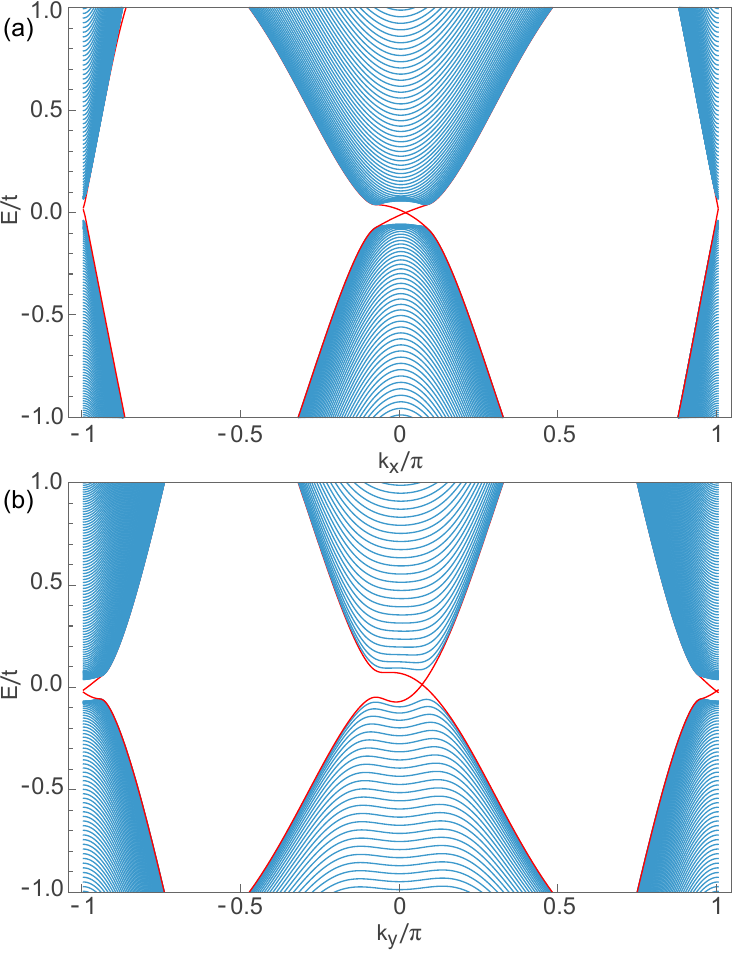}
    \caption{Edge-states in the quantum anomalous Hall (Chern insulator) phase. The parameters are: $t_0=0$, $t_d=0.3t$, $M=2t$, $\lambda_r=0.6t$, $\lambda_s=0.9t$, $\lambda_b=0.2t$.}
    \label{Fig9}
\end{figure}

The narrowness of the $C=-2$ region is a consequence of this
competition. For weak $\alpha_s$, the Berry-curvature hot spots
remain nearly compensated and the phase is trivial. For strong
$\alpha_b$, the bond-induced mass dominates without producing
the required band inversion, again yielding $C=0$. Increasing
$\alpha_s$ eventually drives a topological gap closing and
reopening, but it simultaneously enhances the dispersion and
promotes energy overlap between the valence and conduction
bands. The nontrivial phase therefore occupies only the limited
interval between the topological transition and the loss of the
relevant direct gap.

The magnitude $|C|=2$ can be understood from the multiplicity of
the SOC-free altermagnetic crossings. The square Brillouin zone
contains four symmetry-related Dirac-like nodes, located in
pairs on the $k_x=\pi$ and $k_y=\pi$ boundaries. When the
combined SOC terms gap these nodes with the same topological
mass orientation, each massive cone contributes one half of a
Chern quantum with the same sign, producing a total
$C=-2$. The negative sign follows from the chosen momentum
orientation, magnetic domain, and SOC sign convention; reversing
the magnetic domain reverses the Chern number.

Panel~\ref{Fig5}(e) shows the occupied-band Berry curvature for
$\alpha_s=0$ and $\alpha_b=1.5$. Although the bond SOC produces
finite and strongly momentum-dependent local curvature, the
positive and negative contributions compensate, and the
integrated Chern number remains zero. This illustrates that a
large local Berry curvature does not by itself imply a
topological phase.

Panel~\ref{Fig5}(f) displays the corresponding curvature for
$\alpha_s=1.5$ and $\alpha_b=0.56$. Here the staggered Rashba
coupling removes the cancellation between the symmetry-related
hot spots, and the integrated curvature gives $C=-2$. The
direct gap remains finite, so the Chern number of the lower
two-band manifold is well defined. However, the indirect gap is
negative, meaning that the valence-band maximum and
conduction-band minimum overlap in energy at different
momenta. This state is therefore a Chern band metal rather than
a Chern insulator: it possesses a nontrivial band topology and
can support a pronounced intrinsic Hall and magneto-optical
response, but its dc Hall conductivity is not expected to be
quantized.

Within the parameter window considered here, no finite region
simultaneously exhibits $C\neq0$ and a positive indirect gap except for a larg $M$. For the case of $M>2t$ as shown in Figs.~\ref{Fig5}(c) and (d), we can have Chern insulator with $C=-2$. The edge states corresponding to this case is shwon in Fig.~\ref{Fig9}, confirming $C=2$.
\begin{figure}
    \centering
    \includegraphics[width=0.99\linewidth]{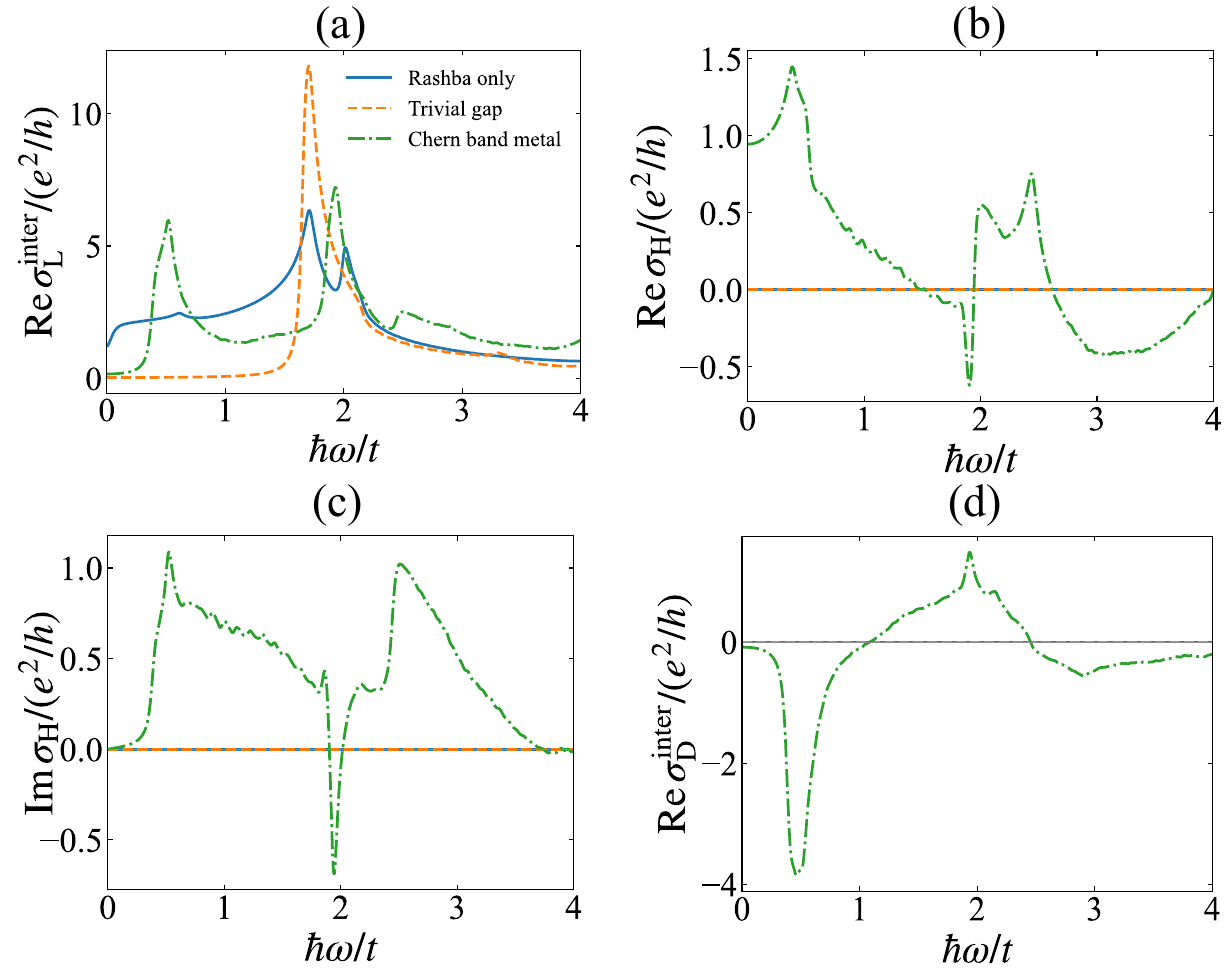}
    \caption{Frequency-dependent optical response for the Rashba-only nodal
phase $(\alpha_s,\alpha_b)=(0,0)$, the trivial gapped phase
$(0,1.5)$, and the Chern band metal $(1.5,0.56)$.
(a) Real part of the interband longitudinal conductivity
$\sigma_{\mathrm L}^{\mathrm{inter}}$.
(b) Real and (c) imaginary parts of the antisymmetric optical Hall
conductivity $\sigma_{\mathrm H}$.
(d) Real part of the interband longitudinal anisotropy
$\sigma_{\mathrm D}^{\mathrm{inter}}$.
The fixed parameters are
$t_d=0.3t$, $M=t$, $t_0=\mu=0$, and $\lambda_R=0.6t$.
The calculation uses $k_BT=0.01t$ and an interband broadening
$\eta=0.03t$. Conductivities are expressed in units of $e^2/h$.
The Drude contribution is excluded from panels (a) and (d).
    }
    \label{Fig6}
\end{figure}

\subsection{Optical and magneto-optical response}

We now examine how the complementary roles of the uniform
Rashba, bond-staggered, and sublattice-staggered SOC terms are
manifested in the frequency-dependent optical response. We
decompose the in-plane conductivity tensor into its average
longitudinal, antisymmetric Hall, and symmetric anisotropic
components,
\begin{equation}
\sigma_{\mathrm L}
=
\frac{\sigma_{xx}+\sigma_{yy}}{2},
\quad
\sigma_{\mathrm H}
=
\frac{\sigma_{xy}-\sigma_{yx}}{2},
\quad
\sigma_{\mathrm D}
=
\frac{\sigma_{xx}-\sigma_{yy}}{2}.
\label{eq:optical_decomposition}
\end{equation}
Here, $\sigma_{\mathrm H}$ controls the intrinsic transverse
magneto-optical response, while $\sigma_{\mathrm D}$ measures
the linear optical anisotropy. The conductivity is evaluated
using the full periodic Hamiltonian and the covariant velocity
operators, including the intra-unit-cell orbital embedding.

Figure~\ref{Fig6} compares three representative phases identified in the
preceding sections: the Rashba-only nodal state
$(\alpha_s,\alpha_b)=(0,0)$, the trivial bond-SOC-gapped
insulator $(0,1.5)$, and the Chern band metal
$(1.5,0.56)$. To expose the finite-frequency transitions,
Figs.~\ref{Fig6}(a) and \ref{Fig6}(d) show only the interband parts of the
longitudinal and anisotropic conductivities. The metallic Drude
contribution is discussed separately below.

Figure~\ref{Fig6}(a) shows that the Rashba-only phase possesses finite
low-frequency absorption, as expected for a nodal spectrum with
no optical threshold. Its dominant interband resonance occurs
near $\hbar\omega\simeq1.7t$. The trivial phase behaves
qualitatively differently: the response is strongly suppressed
below its direct gap,
$\Delta_{\mathrm{dir}}\simeq1.66t$, and rises sharply into a
pronounced absorption peak near $\hbar\omega\simeq1.7t$. The
small residual weight below the gap results from the finite
spectral broadening used in the calculation and vanishes in the
clean limit.

The Chern band metal exhibits a richer two-scale optical
spectrum. A strong low-energy feature appears near
$\hbar\omega\simeq0.5t$, close to the minimum direct gap
$\Delta_{\mathrm{dir}}\simeq0.39t$, and originates from
transitions near the SOC-gapped avoided crossings. A second
prominent structure near $\hbar\omega\simeq1.9t$ arises from
higher-energy transitions involving the reconstructed bands over
a broader region of the Brillouin zone. The common feature near
$1.7t$ in the Rashba-only and trivial phases is therefore not a
direct measure of the low-energy nodal gap, but predominantly a
higher-energy interband joint-density-of-states resonance.

The real and imaginary parts of the optical Hall conductivity
are shown in Figs.~\ref{Fig6}(b) and \ref{Fig6}(c). For both phases with
$\alpha_s=0$, the Hall response vanishes within numerical
precision over the entire frequency range. This confirms that
uniform Rashba SOC and bond-staggered SOC, although capable of
strongly reconstructing and gapping the spectrum, preserve the
antiunitary symmetry that enforces cancellation between
symmetry-related transverse optical transitions.

A qualitatively different response appears when the
sublattice-staggered Rashba term is present. In the Chern band
metal, the symmetry cancellation is removed and
$\sigma_{\mathrm H}(\omega)$ becomes finite. The low-energy
Hall enhancement near $\hbar\omega\simeq0.5t$ coincides with the
first longitudinal absorption peak, demonstrating that the
SOC-gapped crossings carry both large optical matrix elements
and strong band-geometric weight. Further resonant structures
and sign reversals occur near $\hbar\omega\simeq1.9t$ and in the
range $2.4t$--$2.6t$, reflecting competing contributions from
different interband transitions. These finite-frequency sign
changes describe a redistribution of transverse optical
oscillator strength and do not represent changes of the
underlying Chern number.

The static Hall response of the nontrivial phase is finite but
not quantized. Although the isolated lower two-band manifold has
$C=-2$, its indirect gap is negative. Electron and hole pockets
therefore coexist at half filling, and the occupied-state Hall
response contains nonuniversal Fermi-surface contributions. The
state should consequently be regarded as a Chern band metal
rather than a Chern insulator.

Figure~\ref{Fig6}(d) displays the interband longitudinal anisotropy.
Consistent with the surviving fourfold antiunitary symmetry,
$\sigma_{\mathrm D}(\omega)$ vanishes in the two
$\alpha_s=0$ phases. The finite staggered Rashba interaction,
however, lowers the in-plane symmetry and produces pronounced
linear dichroism. At low photon energies,
$\operatorname{Re}\sigma_{\mathrm D}<0$, corresponding to
stronger absorption along $y$ than along $x$. The anisotropy
changes sign at higher frequency and reaches a positive maximum
near the principal $1.9t$ interband resonance. The preferred
absorption axis is therefore frequency dependent.

These results establish the complementary optical functions of
the two staggered SOC terms. The bond-staggered coupling creates
the avoided crossings and associated absorption thresholds,
whereas the sublattice-staggered Rashba interaction removes the
symmetry cancellation of the transverse optical matrix elements.
Their combined action produces simultaneously strong linear
dichroism and a resonant optical Hall response without requiring
a net ferromagnetic moment.

\begin{figure}
    \centering
    \includegraphics[width=0.99\linewidth]{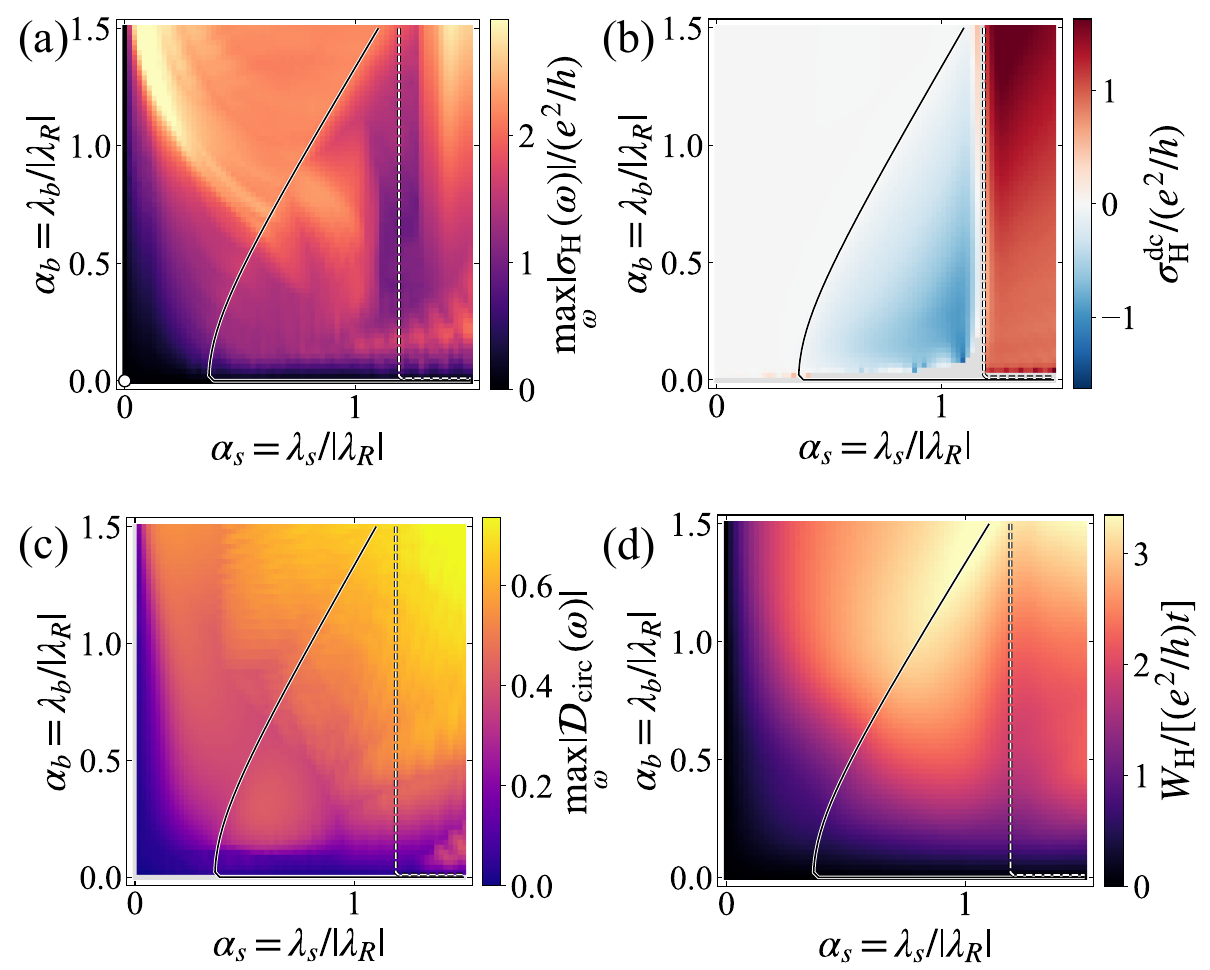}
    \caption{
(a) Maximum magnitude of the complex optical Hall conductivity.
(b) Intrinsic dc Hall conductivity; regions close to a direct-gap
closing are masked.
(c) Maximum normalized circular-dichroism contrast
$\max_{\omega}|\mathcal D_{\mathrm{circ}}(\omega)|$.
(d) Integrated finite-frequency Hall spectral weight
$W_{\mathrm H}=\int d\omega\,
|\sigma_{\mathrm H}(\omega)|$.
The solid black contour denotes
$\Delta_{\mathrm{ind}}=0$ and separates the half-filled
insulating and energy-overlapping metallic regimes. The dashed
white contour encloses the directly gapped manifold with
nonzero Chern number, $C=-2$. Other parameters are the same as Fig.6.
}
    \label{Fig7}
\end{figure}

Having established the frequency-dependent response at selected
parameter points, we now examine how the magneto-optical
properties evolve throughout the
$(\alpha_s,\alpha_b)$ plane.  The quantities displayed in
Fig.~\ref{Fig7} are evaluated over the finite-frequency interval
$\hbar\omega\in[0.15t,3t]$.

To characterize the response without selecting an arbitrary
single photon energy, we introduce the peak optical Hall
amplitude and the integrated Hall spectral weight,
\begin{equation}
\mathcal{S}_{\mathrm H}
=
\max_{\omega}
\left|\sigma_{\mathrm H}(\omega)\right|,
\qquad
W_{\mathrm H}
=
\int d\omega\,
\left|\sigma_{\mathrm H}(\omega)\right|. \nonumber
\label{eq:Hall_optical_measures}
\end{equation}
We also consider the intrinsic dc Hall conductivity
$\sigma_{\mathrm H}^{\mathrm{dc}}$ and the normalized circular
dichroism
\begin{equation}
\mathcal D_{\mathrm{circ}}(\omega)
=
\frac{
\operatorname{Re}\sigma_{+}(\omega)
-
\operatorname{Re}\sigma_{-}(\omega)
}{
\operatorname{Re}\sigma_{+}(\omega)
+
\operatorname{Re}\sigma_{-}(\omega)
}, \nonumber 
\label{eq:circular_dichroism}
\end{equation}
where
$\sigma_{\pm}=\sigma_{\mathrm L}\pm i\sigma_{\mathrm H}$
denote the absorptive conductivities for the two circular
polarizations. For the convention used here,
$\mathcal D_{\mathrm{circ}}
=-\operatorname{Im}\sigma_{\mathrm H}/
\operatorname{Re}\sigma_{\mathrm L}$.
Frequencies with negligible longitudinal absorption are excluded
when determining the maximum dichroism.

Figure~\ref{Fig7}(a) shows the maximum magnitude of the complex optical
Hall conductivity. The response vanishes along the
$\alpha_s=0$ axis, confirming that the uniform Rashba and
bond-staggered SOC terms preserve the antiunitary symmetry that
cancels the transverse optical response. Once
$\alpha_s$ becomes finite, this cancellation is removed and a
broad Hall-active region develops. The strongest resonances occur
when the sublattice-staggered Rashba coupling is combined with an
appreciable bond-staggered coupling. In this regime,
$\lambda_b$ produces avoided crossings with large interband
oscillator strength, while $\lambda_s$ prevents the associated
Hall contributions from cancelling.

An important result of Fig.~\ref{Fig7}(a) is that the largest
finite-frequency Hall response is not confined to the nonzero
Chern region. Strong resonances also appear in topologically
trivial parts of the phase diagram, particularly near regions of
small band separation. This is expected because the optical Hall
conductivity probes frequency-resolved interband matrix elements
and local band geometry rather than only the Brillouin-zone
integral of the Berry curvature. The narrow structures within
the map indicate crossovers between different dominant optical
transitions and should not be interpreted as additional phase
boundaries.

Figure~\ref{Fig7}(b) displays the intrinsic dc Hall conductivity. In the
trivial insulating region, located on the positive-indirect-gap
side of the solid contour, the static Hall response remains
strongly suppressed. Upon entering the energy-overlapping
metallic regime, the redistribution of Berry curvature and the
appearance of electron and hole pockets produce a finite,
nonuniversal Hall conductivity. Its sign changes as the SOC
parameters drive successive band reconstructions.

The dashed contour in Fig.~\ref{Fig7}(b) encloses the directly gapped
two-band manifold with Chern number $C=-2$. In this region the
dc Hall response has the sign expected from the Berry-curvature
convention used in our calculation. Its magnitude, however, is
not quantized because the indirect gap remains negative. The
phase is therefore a Chern band metal rather than a Chern
insulator. Regions sufficiently close to a direct-gap closing
are omitted from panel~(b), since the clean dc interband response
becomes sharply momentum dependent and is not numerically stable
at the transition itself.

Figure~\ref{Fig7}(c) shows the maximum circular-dichroism contrast. A
substantial difference between right- and left-circular
absorption develops over a broad region of the parameter plane
and reaches values of order unity without exceeding the
passivity bound. As for the peak Hall conductivity, strong
circular dichroism is not restricted to the Chern region. It is
instead enhanced wherever SOC-induced avoided crossings combine
large optical matrix elements with an asymmetric distribution of
transverse interband weight. Since the maximum is taken
independently at each parameter point, panel~(c) represents the
largest achievable contrast within the selected photon-energy
window rather than the response at a common fixed frequency.

The integrated Hall spectral weight in Fig.~\ref{Fig7}(d) provides a
smoother measure of the overall transverse optical activity.
Unlike the height of an individual resonance, $W_{\mathrm H}$ is
less sensitive to the precise resonance width and frequency-grid
resolution. It vanishes when $\alpha_s=0$ and grows rapidly when
both staggered SOC channels are present, reaching its largest
values for intermediate-to-strong $\alpha_s$ and $\alpha_b$.
The broad maximum confirms that the enhanced optical Hall effect
is not caused by a single isolated spectral peak but is
distributed over a finite range of interband transitions.

The comparison of the four panels separates the roles of band
topology and finite-frequency optical geometry. The narrow
$C=-2$ region controls the topological character of the isolated
lower-band manifold and strongly influences the static Hall
response. By contrast, resonant Hall conductivity, circular
dichroism, and integrated Hall spectral weight remain large over
a substantially wider region that includes topologically trivial
states. Consequently, a strong magneto-optical signal does not
by itself establish a nonzero Chern number, but instead provides
a sensitive probe of the combined nodal-gap formation and
symmetry breaking generated by $\lambda_b$ and $\lambda_s$.

\begin{figure}
    \centering
    \includegraphics[width=0.99\linewidth]{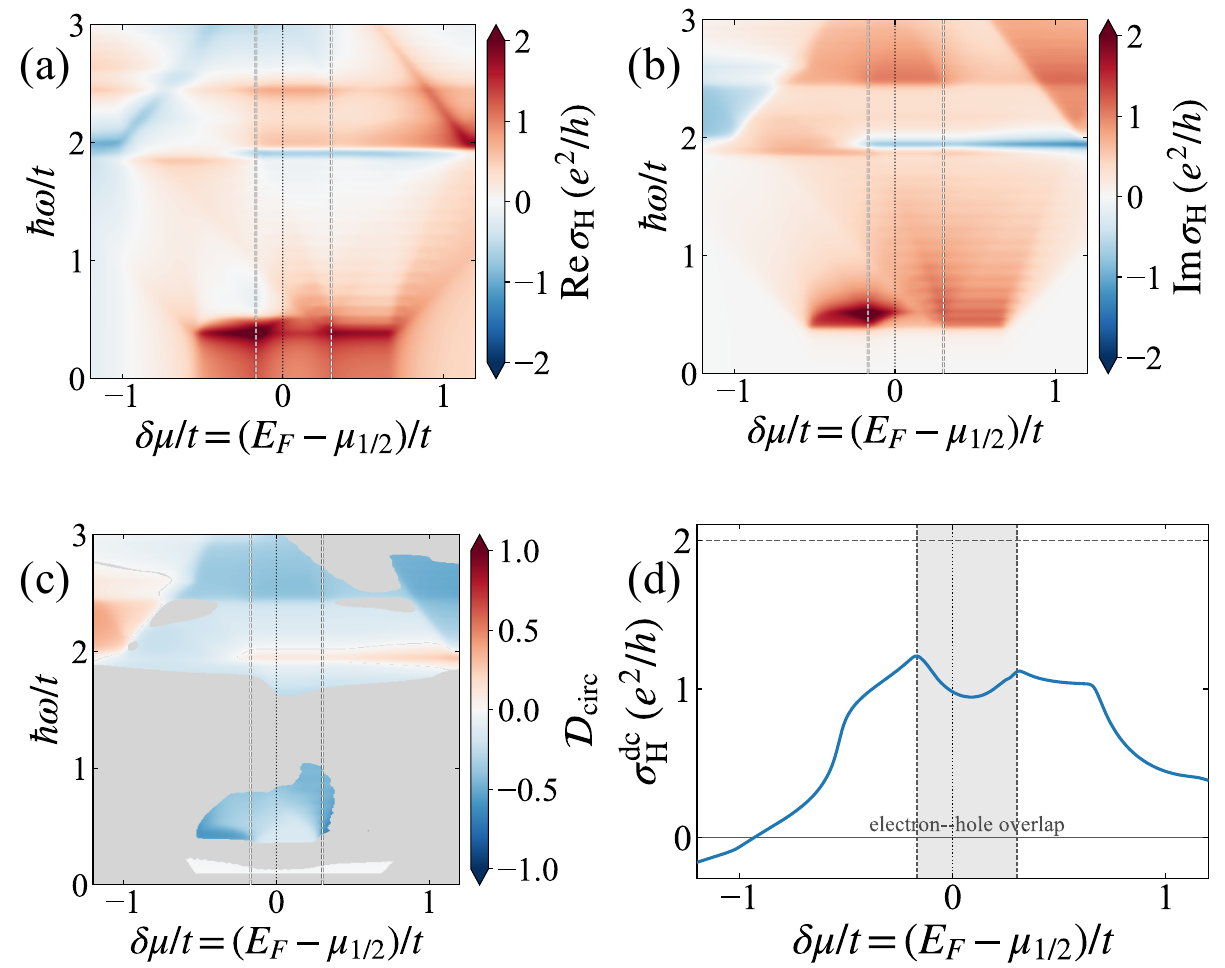}
    \caption{
(a) Real and (b) imaginary parts of the optical Hall
conductivity.
(c) Normalized circular-dichroism contrast.
Gray regions indicate frequencies at which the longitudinal
absorption or Hall response is too small for a reliable
normalized contrast.
(d) Intrinsic dc Hall conductivity as a function of chemical
potential. The shaded interval denotes the electron--hole
coexistence region associated with the negative indirect gap,
and the horizontal dashed line marks the quantized
Chern-insulator reference
$-C e^2/h=2e^2/h$.
In all panels, the dotted vertical line denotes exact half
filling and the two vertical dashed lines mark the conduction-band
minimum and valence-band maximum that bound the electron--hole
overlap window.
The remaining parameters are
$t_d=0.3t$, $M=t$, $t_0=0$, and $\lambda_R=0.6t$, with
$k_BT=0.01t$ and optical broadening $\eta=0.03t$.
}
    \label{Fig8}
\end{figure}

We finally investigate the dependence of the magneto-optical
response on carrier occupation. We focus on the representative
Chern band metal at
$(\alpha_s,\alpha_b)=(1.5,0.56)$, for which the isolated lower
two-band manifold carries $C=-2$, while the negative indirect
gap produces coexisting electron and hole pockets. To identify
half filling unambiguously, the gate-controlled chemical
potential is measured relative to the value $\mu_{1/2}$ that
satisfies $n(\mu_{1/2})=2$, and
$\delta\mu=E_F-\mu_{1/2}.
$The dotted vertical line in Fig.~8 denotes $\delta\mu=0$, while
the two dashed lines delimit the electron--hole overlap window.

Figures~\ref{Fig8}(a) and \ref{Fig8}(b) show the real and imaginary parts of the
optical Hall conductivity as functions of photon energy and
chemical potential. A pronounced low-energy response appears
near $\hbar\omega\simeq0.4$--$0.5t$. This feature originates
from transitions across the small SOC-induced direct gaps near
the reconstructed altermagnetic crossings. It is strongest
close to half filling, where both the relevant occupied and
unoccupied states are available, and is progressively suppressed
upon electron or hole doping by Pauli blocking.

Additional Hall structures occur near
$\hbar\omega\simeq1.9t$ and in the range
$2.4t$--$2.6t$. These higher-energy features correspond to
transitions involving more remote portions of the reconstructed
bands and are therefore less sensitive to moderate changes of
the Fermi level. Their positions agree with the resonances found
in the fixed-filling spectra of Fig.~\ref{Fig6}. The alternating positive
and negative regions in
$\operatorname{Re}\sigma_{\mathrm H}$ and
$\operatorname{Im}\sigma_{\mathrm H}$ reflect competition
between different interband channels and the redistribution of
Berry-curvature-weighted optical matrix elements. They should
not be interpreted as changes of the Chern number, since the
Hamiltonian and the directly isolated band manifold remain
unchanged throughout this gate scan.

Figure~\ref{Fig8}(c) displays the normalized circular-dichroism contrast,
\begin{equation}
\mathcal D_{\mathrm{circ}}
=
\frac{
\operatorname{Re}\sigma_{+}
-
\operatorname{Re}\sigma_{-}
}{
\operatorname{Re}\sigma_{+}
+
\operatorname{Re}\sigma_{-}
}
=
-
\frac{
\operatorname{Im}\sigma_{\mathrm H}
}{
\operatorname{Re}\sigma_{\mathrm L}
}.\nonumber
\end{equation}
A sizable contrast develops around the same low- and
high-energy interband resonances that dominate the optical Hall
conductivity. Its sign reversals demonstrate that the preferred
circular polarization can be switched by varying either the
photon energy or the carrier density. Gray regions correspond to
frequencies at which the longitudinal absorption or Hall signal
is too small for a reliable normalized dichroism to be assigned.

The intrinsic dc Hall conductivity is shown in Fig.~\ref{Fig8}(d). Near
half filling, $\sigma_{\mathrm H}^{\mathrm{dc}}$ remains of order
$e^2/h$ and exhibits local maxima close to the boundaries of the
electron--hole overlap region, where the Fermi level passes
through band extrema carrying enhanced Berry curvature. At
larger electron or hole doping, the response decreases and may
change sign as additional momentum-space regions with opposite
Berry curvature become occupied.

The dashed horizontal line in Fig.~\ref{Fig8}(d) denotes the quantized
Chern-insulator reference
$-C e^2/h=2e^2/h$ for the sign convention used here. The
calculated Hall conductivity does not approach this value because
the indirect gap is negative and the Fermi level intersects
electron and hole pockets. Thus, although the isolated lower
two-band manifold retains $C=-2$, the measurable dc Hall
conductivity is nonquantized and varies continuously with
carrier occupation.

Figure~\ref{Fig8} therefore distinguishes the topology of an isolated
band manifold from the transport response of a metallic filling.
The Chern number remains fixed, whereas both the optical and dc
Hall conductivities can be strongly modulated through Pauli
blocking and the occupation of Berry-curvature hot spots. This
gate tunability provides an experimentally accessible means of
controlling the magnitude and sign of the magneto-optical
response without altering the magnetic order or the underlying
SOC parameters.

\section{Conclusions}

We have developed a minimal, strictly periodic four-band model for a
two-dimensional $d$-wave altermagnet and used it to clarify how
different spin--orbit-coupling mechanisms control the electronic,
topological, and magneto-optical properties of the system. In the
absence of SOC, the combination of staggered exchange and
sublattice-dependent $d$-wave hopping produces momentum-dependent spin
splitting with vanishing net magnetization. The resulting spin-up and
spin-down band crossings lie on orthogonal Brillouin-zone
boundaries and are related by the altermagnetic fourfold spin-group
symmetry. Their appearance requires $t_d\geq M/4$, although
altermagnetic spin splitting itself is already present for any
nonzero $t_d$ and $M$ away from the symmetry-enforced nodal lines.

The three SOC terms considered here play distinct and complementary
roles. Uniform Rashba SOC mixes the two spin sectors and reconstructs
the spin texture, but it preserves the antiunitary symmetry that
relates opposite transverse responses and therefore does not by itself
generate a finite integrated Hall conductivity. The
sublattice-staggered Rashba interaction breaks this cancellation and
activates optical Hall response and circular dichroism, while generally
leaving the boundary crossings ungapped. In contrast, the
bond-staggered SOC preserves $S_z$ but acts as a mass for the
Dirac points, producing a directly gapped spectrum while remaining
topologically trivial when the Hall-forbidding symmetry is intact.
A finite transverse response and a robust nodal mass therefore require
the combined action of the staggered Rashba and bond-staggered SOC
channels.

The corresponding gap maps reveal that a positive direct gap occupies
a broad region of the staggered-SOC parameter space, whereas a positive
indirect gap is restricted to a smaller domain. This distinction is
essential: the Chern number of the isolated lower two-band manifold is
well defined whenever the direct gap remains open, but a quantized Hall
state additionally requires a positive indirect gap. Within the
parameter range examined here, a narrow nontrivial region with
$C=-2$ emerges from an SOC-driven band inversion. The magnitude
$|C|=2$ reflects the combined contributions of the symmetry-related
massive Dirac-like crossings.
For sufficiently large sublattice magnetization $M$, the non-trivial region becomes a Chern insulator phase. On the other hand, for lower $M$, for instance $M=1t$ used
in the optical calculations, the indirect gap remains
negative. The nontrivial state is therefore a Chern band metal rather
than a Chern insulator, and its dc Hall conductivity is finite but not
quantized.

The optical response reflects this distinction between topology and
finite-frequency band geometry. Frequency-resolved Kubo calculations
show that the Rashba-only and bond-SOC-gapped phases have vanishing
antisymmetric optical Hall conductivity, in agreement with symmetry.
Once the sublattice-staggered Rashba term is introduced, pronounced
Hall resonances and linear and circular dichroism appear near the
SOC-induced avoided crossings. The strongest low-energy response is
associated with transitions across the small direct gaps near the
reconstructed altermagnetic nodes, while higher-energy structures arise
from transitions involving more remote parts of the band structure.

A central result is that strong finite-frequency magneto-optical
activity is not confined to the nonzero-Chern region. Large optical
Hall conductivity, circular dichroism, and integrated Hall spectral
weight extend over a much broader parameter range that includes
topologically trivial states. The bond-staggered SOC controls the
strength of the avoided-crossing transitions, whereas the
sublattice-staggered Rashba term removes the symmetry cancellation
between their transverse contributions. The Chern topology is instead
most clearly reflected in the structure of the intrinsic dc Hall
response and in the gap closing and reopening of the isolated
two-band manifold.

We further showed that the magneto-optical response is strongly tunable
by carrier occupation. In the $C=-2$ Chern band metal, changing the
chemical potential modifies the available interband transitions
through Pauli blocking and changes which Berry-curvature hot spots are
occupied. As a result, both the optical and dc Hall conductivities can
be continuously enhanced, suppressed, or reversed in sign without
altering the magnetic order or the SOC parameters. The preferred
circular polarization can likewise be switched by tuning either photon
energy or carrier density. This gate control provides a direct way to
separate the fixed topology of the band manifold from the
nonuniversal transport response of a metallic filling.

Our results establish a unified microscopic picture in which nodal
mass generation, symmetry breaking, band topology, and optical
activity originate from different but cooperating SOC channels. They
also show that a large magneto-optical signal should not be interpreted
as evidence of a nonzero Chern number by itself. Instead, the combined
analysis of direct and indirect gaps, dc Hall response,
frequency-dependent Hall conductivity, and circular dichroism is
required to distinguish a trivial Hall-active metal, a Chern band
metal, and a true Chern insulator.

The model provides a useful framework for interface-engineered
two-dimensional altermagnets, where uniform and staggered SOC terms can
be tuned independently through substrate choice, structural registry,
electric fields, and carrier doping. Extensions including
material-specific hopping parameters, disorder, interactions, finite
thickness, and the electromagnetic environment of a realistic
substrate will be important for quantitative comparison with
experiment. Nevertheless, the present results demonstrate that
compensated $d$-wave altermagnets can host SOC-controlled topology and
strong, gate-tunable magneto-optical response without requiring a net
ferromagnetic moment.

\begin{acknowledgments}
M.B.T. acknowledges the funding support by  Narodowa Agencja Wymiany Akademickiej (NAWA) under the ULAM program with project number BPN/ULM/2025/1/00156/U/00001.
M.B.T. acknowledges the funding support by Iran National Science Foundation (INSF) under project No.4043973. This research was supported by the Foundation for Polish Science project “MagTop” no. FENG.02.01-IP.05-0028/23 co-financed by the European Union from the funds of Priority 2 of the European Funds for a Smart Economy Program 2021–2027 (FENG). We further acknowledge access to the computing facilities of the Interdisciplinary Center of Modeling at the University of Warsaw, Grant g91-1418, g91-1419, g96-1808, g96-1809 and g103-2540 for the availability of high-performance computing resources and support. We acknowledge the access to the computing facilities of the Poznan Supercomputing and Networking Center, Grants No. pl0267-01, pl0365-01, pl0471-01, and pl0807.
\end{acknowledgments}

\medskip

\appendix

\bibliography{Referencesv1}
\end{document}